\documentclass[english]{article}
\usepackage[T1]{fontenc}
\usepackage[latin9]{inputenc}
\usepackage{geometry}
\usepackage{amsmath}
\usepackage{amssymb}
\usepackage{graphicx}
\usepackage{setspace}
\usepackage{esint}
\makeatletter

\providecommand{\tabularnewline}{\\}

\makeatother

\usepackage{babel}
\begin{document}
\begin{doublespace}

\title{Controlled Finite Momentum Pairing and Spatially Varying Order Parameter
in Proximitized HgTe Quantum Wells}
\end{doublespace}

\begin{singlespace}

\author{\noindent Sean Hart$^{\dagger1}$, Hechen Ren$^{\dagger1}$, Michael
Kosowsky$^{1}$,\\
 Gilad Ben-Shach$^{1}$, Philipp Leubner$^{2}$, Christoph Brüne$^{2}$,
Hartmut Buhmann$^{2}$, \\
 Laurens W. Molenkamp$^{2}$, Bertrand I. Halperin$^{1}$, Amir Yacoby$^{1}$}
\end{singlespace}

\begin{singlespace}

\date{\noindent $^{1}$Department of Physics, Harvard University, Cambridge,
MA, USA \\
$^{2}$ Physikalisches Institut (EP3), Universität Würzburg, 97074
Würzburg, Germany\\
$^{\dagger}$These authors contributed equally to this work}
\end{singlespace}
\maketitle
\begin{abstract}
Conventional $s$-wave superconductivity is understood to arise from
singlet pairing of electrons with opposite Fermi momenta, forming
Cooper pairs whose net momentum is zero \cite{Bardeen1957}. Several
recent studies have focused on structures where such conventional
$s$-wave superconductors are coupled to systems with an unusual configuration
of electronic spin and momentum at the Fermi surface. Under these
conditions, the nature of the paired state can be modified and the
system may even undergo a topological phase transition \cite{Fu2008,Sau2010}.
Here we present measurements and theoretical calculations of several
HgTe quantum wells coupled to either aluminum or niobium superconductors
and subject to a magnetic field in the plane of the quantum well.
By studying the oscillatory response of Josephson interference to
the magnitude of the in-plane magnetic field, we find that the induced
pairing within the quantum well is spatially varying. Cooper pairs
acquire a tunable momentum that grows with magnetic field strength,
directly reflecting the response of the spin-dependent Fermi surfaces
to the in-plane magnetic field. In addition, in the regime of high
electron density, nodes in the induced superconductivity evolve with
the electron density in agreement with our model based on the Hamiltonian
of Bernevig, Hughes, and Zhang \cite{Bernevig2006_Science}. This
agreement allows us to quantitatively extract the value of $\tilde{g}/v_{F}$,
where $\tilde{g}$ is the effective g-factor and $v_{F}$ is the Fermi
velocity. However, at low density our measurements do not agree with
our model in detail. Our new understanding of the interplay between
spin physics and superconductivity introduces a way to spatially engineer
the order parameter, as well as a general framework within which to
investigate electronic spin texture at the Fermi surface of materials.
\end{abstract}
Below a critical temperature and magnetic field, certain materials
undergo a phase transition to the superconducting state. Macroscopically
identified through effects such as zero resistivity and the Meissner
effect \cite{Meissner1933}, superconductors may further be understood
microscopically as arising due to pairing of electrons occupying opposite
points on the Fermi surface and having opposite spin. Within a conventional
setting this interaction results in Cooper pairs with zero net momentum.
However, in certain materials the presence of both magnetic order
and superconductivity can lead to intrinsically nonzero pairing momentum
as the system enters the Fulde-Ferrell-Larkin-Ovchinnikov (FFLO) state
\cite{FF1964,LO1965}. Studies of both CeCoIn$_{5}$ and $\kappa$-(BEDT-TTF)$_{2}$Cu(NCS)$_{2}$
under large external magnetic fields found evidence for coupled magnetic
order and superconductivity, although in each material the field strength
needed was in excess of 10 T \cite{Kenzelmann2008,Mayaffre2014}.

Exotic superconductivity has recently come under additional investigation
through the goal of combining $s-$wave superconductors with materials
whose properties are rarely found among the conventional superconductors.
For example, inducing the singlet pairing of an $s$-wave superconductor
into a material with strong spin-orbit coupling and reduced dimensionality
has been recently considered as a viable platform within which to
achieve triplet pairing \cite{Reeg2015} and topological superconductivity
\cite{Fu2008,Sau2010}, or to engineer a Josephson $\phi_{0}$-junction
\cite{Yokoyama2014,Dolcini2015}. Moreover, when a ferromagnetic layer
is sandwiched by two superconductors, pairs traversing the junction
acquire momentum due to the exchange field within the ferromagnet
\cite{Buzdin1982,Demler1997}. Measurements of critical current oscillations
in such superconductor-ferromagnet-superconductor (SFS) junctions
have provided evidence for this nonzero pairing momentum \cite{Kontos2002,Sellier2003,Frolov2004},
although the magnitude of the momentum was effectively untunable due
to the typically large exchange fields. 

Here we report on coupling between superconducting leads and a two-dimensional
electron system realized within HgTe/HgCdTe heterostructures in the
inverted regime. Due to the interplay between superconductivity, band
structure, and the applied magnetic field, we find that the order
parameter has an oscillatory component derived from the finite momentum
of paired electrons, and that this momentum can be continuously tuned
between conventional and unconventional regimes. Our use of only relatively
small external magnetic fields ($\leq4$ T) and micron-scale device
dimensions introduces a new regime in the exploration of the interplay
between superconductivity and spin physics.

To study the effect of magnetic field and band structure on electron
pairing, we place two superconducting leads on opposite boundaries
of a rectangular section of quantum well. Devices were fabricated
at varying angles with respect to the cleavage edges of the crystal
(the $[110]$ and $[1\bar{1}0]$ axes). The angular alignment corresponds
to a rotation angle $\theta$ with respect to the principal crystal
axis $[100]$, with $\theta$ defined modulo $\pi/2$ (Figure 1a,
see Supplementary Information I). The width $W$ between the two leads
is 800 nm and the length $L$ of the resulting Josephson junction
is 4 microns. We study the influence of either niobium or aluminum
superconductors by applying a small AC current bias between the two
leads while measuring the resultant AC voltage \cite{Oostinga2013}.
The aluminum thickness is 15 nm in order to sustain superconductivity
in moderate parallel magnetic fields (Supplementary Information II)
\cite{Meservey1971}, while the niobium thickness is 130 nm. Josephson
interference is generated by application of small (up to $\sim10$
mT) magnetic fields in the $z$ direction \cite{Tinkham2004}. Throughout,
the in-plane coordinate axes are referred to as $x$ and $y$, respectively
oriented perpendicular and parallel to the supercurrent flow between
leads. The addition of a normal metal topgate allows us to tune the
electron density within the junction, thereby providing a means to
study superconductivity from the quantum spin Hall regime \cite{Konig2007,Hart2014}
into the electron-doped regime. In the regime of high electron density
and with no parallel magnetic field, junctions show Josephson interference
consistent with uniform supercurrent transport through the bulk of
the quantum well, shown in Figure 1b for a device with aluminum leads.

We primarily study differential resistance with zero applied DC current,
due to the efficiency of such measurements in illuminating the structure
of the interference pattern. Lower resistance relative to the normal
device resistance typically corresponds to elevated critical current
(Supplementary Information III). In an aluminum-based junction, in
the electron-doped regime and with angle $\theta=\pi/4$, increasing
the magnetic field in the $x$ direction strongly modulates the Josephson
interference (Figure 1c). Two distinct regions of decreased resistance
are separated by a nodal field of approximately $B_{x}=1.1$ T, corresponding
to the suppression of induced superconductivity. At each value of
the parallel field, we extract the minimum junction resistance as
a measure of the strength of superconductivity at that particular
field. Plotting these minimum junction resistances highlights the
oscillatory effect of parallel field on superconductivity, with the
nodal field marked by an arrow (Figure 1d). The suppression of superconductivity
at the nodal field directly results from the finite momentum of induced
Cooper pairs.

In an aluminum-based junction oriented with $\theta=\pi/2$, a similar
modulation of superconductivity occurs as the parallel field $B_{x}$
grows (Figure 1e, see Supplementary Information IV-V). Although the
aluminum leads can sustain superconductivity up to 1.75 T, we only
measure this device up to $B_{x}=1$ T due to constraints on the range
of our vector magnet (see Supplementary Information I).

The resistance of a device with niobium leads and $\theta=\pi/4$
is similarly modulated upon application of a parallel field, with
multiple nodes visible as $B_{x}$ increases to 4 T (Figure 1f). A
more detailed measurement highlights the presence of three distinct
regions of decreased resistance, separated by bands of high resistance
occurring near $B_{x}=0.9$ T and $B_{x}=2.7$ T (Figure 1g). We again
extract the minimum junction resistance at each particular parallel
field value, demonstrating the oscillatory effect of parallel field
on superconductivity (Figure 1h). Nodes of the oscillation, marked
by arrows, correspond to local maxima in the overall junction resistance.

Despite the differences in fabrication of our devices, the nodal structure
is both robust and occurs at nearly the same parallel field magnitudes.
These observations suggest that the induced pairing momentum originates
in the heterostructures and not the bulk superconductors, and is insensitive
to details of the crystal orientation. Since superconductivity arises
from pairing of electrons with opposing spins and momenta, it is therefore
necessary to examine the nature of both Zeeman coupling and spin-orbit
coupling within the quantum well. 

We model our devices by considering first the quantum well region
in the absence of the superconductors, for which a four-band theoretical
Hamiltonian $H_{1}$ was proposed as a way to describe the topology
of the band structure \cite{Bernevig2006_Science}. We adopt a version
of this model to include both the external magnetic field and possible
contributions from spin-orbit coupling \cite{Rothe2010,Weithofer2013,Konig2008}.
The key prediction of the band structure modeling is that the Zeeman
coupling from the external field $B_{x}$ modifies the Fermi surfaces
in a manner which depends on the nature of the spin-orbit coupling
(Supplementary Information VI). As a consequence, the induced superconducting
order parameter is expected to oscillate in space, due to a pairing
momentum shift with magnitude of order $\hbar\Delta k\approx\tilde{g}\mu_{B}B_{x}/v_{F}$,
whose orientation also depends on the spin-orbit coupling. Here $\tilde{g}$
is the in-plane g-factor, and $v_{F}$ is the Fermi velocity.

To theoretically investigate the proximity effect in our quantum wells,
we consider a geometry in which the two-dimensional electron gas (2DEG),
assumed to have uniform electron density, is contacted by a pair of
superconducting leads with a controlled phase difference between them,
and we seek to calculate the maximum supercurrent that can be carried
between the strips (Figure 3a, see Supplementary Information VII-X
for details not presented in the main text). We assume a Hamiltonian
$H=H_{1}+H_{2}$, where $H_{2}$ is the coupling between the superconductors
and the 2DEG, described by a pairing Hamiltonian of the form

\begin{equation}
H_{2}=-\int dxdy\left[\Delta(x,y)\Psi^{\dagger}(x,y)+\Delta^{*}(x,y)\Psi(x,y)\right].
\end{equation}

\noindent Here $\Psi(x,y)\equiv\psi_{\uparrow}(x,y)\psi_{\downarrow}(x,y)$
is an operator which annihilates a singlet pair of electrons in the
2DEG at the point $(x,y)$, while the pair potential $\Delta(x,y)$
is a complex number that depends on the phase of the superconductor
and the tunneling amplitude at that point.

We assume that the contacts between the 2DEG and the superconductors
occur at the edges of the superconductors, located at $y=0$ and $y=W$,
so that we may write

\begin{equation}
\Delta(x,y)=\lambda_{1}(x)\delta(y)+\lambda_{2}(x)\delta(y-W)\equiv\Delta_{1}(x,y)+\Delta_{2}(x,y),
\end{equation}

\noindent with $-L/2<x<L/2$. We assume that the magnitude of the
coupling is constant along each lead, but the phase will vary if there
is a perpendicular magnetic field $B_{z}\neq0.$ We choose a gauge
where the vector potential points in the $x$ direction, with $A_{x}=-B_{z}(y-W/2)$,
so that the vector potential vanishes along the midline of the 2DEG.
If the superconducting strips have identical widths $W_{SC}$, then
the couplings $\lambda_{j}$ will have the form

\begin{equation}
\lambda_{j}(x)=\left|\lambda_{j}\right|e^{2\pi i\phi_{j}(x)},
\end{equation}

\begin{equation}
\phi_{j}(x)=\phi_{j}(0)+\frac{(-1)^{j-1}xB_{z}(W+W_{sc})}{2\Phi_{0}},
\end{equation}

\noindent with $j=1,2$.

To lowest order in the pairing Hamiltonian $H_{2}$, the portion of
the total energy that depends on the phase difference between the
two superconducting leads can be written in the form:

\begin{equation}
E=-\int dx_{2}\left[\lambda_{2}^{*}(x_{2})\left\langle \Psi(x_{2},W)\right\rangle _{1}+c.c.\right],
\end{equation}

\noindent where $\left\langle \Psi(x,y)\right\rangle _{1}$ is the
order parameter at point $(x,y)$ induced by the superconductor $j=1$.
In turn, this may be written in the form

\begin{equation}
\left\langle \Psi(x,y)\right\rangle _{1}=\int dx_{1}\lambda_{1}(x_{1})F(x,x_{1},y),
\end{equation}

\noindent where $F$ is the propagator from point $(x_{1},0)$ to
point $(x,y)$ for an induced Cooper pair. Depending upon the relative
magnitudes of spin-orbit coupling and the Zeeman coupling, the propagator
$F$ may take various forms (see Supplementary Information IX). In
the limit where either structural inversion asymmetry (SIA) or bulk
inversion asymmetry (BIA) is strong compared to the Zeeman coupling,
the pair momentum shift orientation is independent of position on
the Fermi surface. The shift occurs along an angle $\alpha$ with
respect to the $x$ axis, and the propagator is

\begin{equation}
F(x,x_{1},y)=\frac{k_{F}}{8\pi{}^{2}v_{F}}\cdot\frac{e^{i\gamma}+e^{-i\gamma}}{(x-x_{1})^{2}+y^{2}},\gamma=\Delta k\left(\sin(\alpha)y+\cos(\alpha)(x-x_{1})\right).
\end{equation}

When SIA dominates the spin-orbit coupling, a magnetic field $B_{x}$
induces pairing momentum in the $y$ direction, and the order parameter
also oscillates in the $y$ direction (Figure 2a). When $\Delta kW=\pi/2$,
the first node of the oscillation coincides with the line $y=W$ corresponding
to the width of the junction. Increasing the parallel field so that
$\Delta kW=3\pi/2$ leads to coincidence of the second node and the
junction width (Figure 2b).

If BIA instead dominates the spin-orbit coupling, when $\theta=0$
the parallel magnetic field induces order parameter oscillations in
the $x$ direction (Figure 2c). These oscillations arise due the finite
length of the Josephson junction, with amplitudes that are largest
near the ends of the mesa. In contrast to the limit of large SIA,
with dominant BIA the nodes of the order parameter never coincide
with the junction width. Oscillations in the order parameter instead
occur with greater frequency along the $x$ direction as the magnetic
field increases.

Finally, when the Zeeman coupling dominates the spin-orbit coupling,
the pair momentum shift magnitude is isotropic in-plane, but the orientation
lies parallel to the direction of Cooper pair propagation. In this
limit the propagator is

\begin{equation}
F(x,x_{1},y)=\frac{k_{F}}{8\pi{}^{2}v_{F}}\cdot\frac{e^{i\gamma}+e^{-i\gamma}}{(x-x_{1})^{2}+y^{2}},\gamma=\Delta k\sqrt{(x-x_{1})^{2}+y^{2}}.
\end{equation}

Here the induced order parameter oscillates along both the $x$ and
$y$ directions (Figure 2d). Although the shape of the order parameter
resembles the limit of strong SIA, the possibility to oscillate in
all in-plane directions prevents a node from forming along a line
of constant $y.$

We can link the order parameter oscillations to the Josephson energy
$E$ by integrating over the second superconducting lead at position
$y=W$, as in equation (5). By then differentiating with respect to
the phase difference $\phi_{2}(0)-\phi_{1}(0)$ we find the current-phase
relation of the junction, which is then maximized with respect to
the phase difference to obtain the critical current. 

When SIA dominates the spin-orbit coupling, the critical current periodically
disappears when the nodal condition $\Delta kW=(2n+1)\pi/2$ is satisfied
(Figure 3b). These nodal points in the critical current correspond
to spatial nodes of the induced order parameter coinciding with the
positions of the leads, so that the supercurrent is completely suppressed.
Microscopically, these oscillations of the order parameter correspond
to finite momentum pairing of electrons, as diagrammed in the inset
of Figure 3b. In the limit of strong SIA, the Fermi surfaces oppositely
shift in the $y$ direction, so that Cooper pairs form internally
to each surface with finite wavevector $\Delta k\hat{y}$.

The predicted interference with strong SIA resembles the nodal pattern
we observe experimentally. However, in niobium-based devices we also
observe that superconductivity weakens and that nodes occur over an
increasingly large range of parallel field strengths as the parallel
field increases. To model this behavior, we consider the effect of
structural imperfections in the interface between the superconducting
leads and the quantum well. These would introduce random fluctuations
in the direction of the in-plane magnetic field at each interface,
leading to a random component of the pairing momentum in the $x$
direction that grows linearly with the in-plane field. Hence, we introduce
a random phase $\chi\propto\left(R_{1}(x_{1})-R_{2}(x_{2})\right)B_{x}$,
where the random variables $R_{1}(x_{1})$ and $R_{2}(x_{2})$ correspond
to fluctuations in the direction of the parallel field at each interface
(see Supplement XI for details). With this randomness, the calculated
critical currents diminish in magnitude as the in-plane field increases,
and nodes occur over a larger range of in-plane field, in agreement
with our experimental observation (Figure 3c).

Considering, instead, BIA as the dominant source of spin-orbit coupling,
when the junction is aligned to the $[100]$ or $[010]$ crystal direction,
the order parameter oscillates in the $x$ direction. This oscillation
corresponds to shifting of the Fermi surfaces oppositely along $x$,
so that Cooper pairs form internally to each surface with wavevector
$\Delta k\hat{x}$ (Figure 3d). Since the real-space supercurrent
density and the Josephson critical current can be regarded as Fourier
conjugates \cite{Dynes1971}, this pairing momentum results in finite
weight of the interference at a particular magnitude of $B_{z}$ that
grows linearly with the parallel field, forming a `V' shape. In our
measurements of the device oriented with $\theta=\pi/2$, this splitting
would be seen in the limit of strong BIA, but is not observed experimentally
(Figure 1e, Supplementary Information V). Additionally, when junctions
are fabricated at an angle $\theta=\pi/4$, with strong BIA the behavior
is expected to shift from that shown in Figure 3d to the nodal structure
in Figure 3b. Since we instead observe behavior that does not depend
on the crystal orientation, we conclude that BIA in our heterostructure
is relatively weak. This conclusion agrees with a previous measurement
of Shubnikov-de Haas oscillations in a HgTe quantum well, which was
found to be consistent with strong SIA and weak BIA \cite{Gui2004}.

In the limit of overall weak spin-orbit coupling, the order parameter
oscillates in both in-plane directions. Zeeman coupling at finite
values of $B_{x}$ leads to two concentric Fermi surfaces with opposite
spin polarization, so that pairing occurs between surfaces with momentum
in all in-plane directions (Figure 3e). Increasing the parallel magnetic
field causes the interference to both spread in $B_{z}$ and periodically
oscillate, a hybrid of the two above cases. Characteristically, at
each node the two interference fringes adjacent to the central fringe
combine to form the subsequent central fringe, a direct result of
the inability to form nodes in the order parameter along lines of
constant $y$. Although it is possible that this behavior is present
in the device with niobium leads, the nodal pattern is more consistent
with strong SIA with aluminum leads at high density. In the limit
of overall weak spin-orbit coupling, with the random phase $\chi$
the modeled interference successfully reproduces many aspects of the
behavior observed in the niobium device, but is still inconsistent
with the aluminum devices (Figure 3f). 

As an additional study into the nature of electron pairing momentum,
we explore the evolution of the minimum junction resistance at different
parallel field $B_{x}$ values, while energizing the global topgate
to modify the bulk electron density. Devices used for these measurements
were aligned such that $\theta=\pi/4$, corresponding to the devices
of Figure 1c,d,f-h. At the most positive gate voltage, as the magnetic
field is increased the niobium device displays the narrow node of
increased resistance near $B_{x}=0.9$ T (Figure 4a). As before, an
additional wider node is present near $B_{x}=2.7$ T. When the top
gate voltage is lowered to $-5$ V, the field magnitude of the lower
node increases, first slowly and then more rapidly. In the device
with aluminum leads, a similar nodal structure is observed, with the
magnitude of the node weakly increasing as the top gate voltage is
lowered from $0.5$ V to $0$ V (Figure 4b). 

The dependence of the node magnitude on density can be calculated
within the framework of our model based on BHZ theory, here assuming
the presence of SIA due to a perpendicular electric field equal to
10 mV/nm (Supplementary Information VI). Since the magnitude of the
induced Cooper pair momentum is approximately $\hbar\Delta k\approx\tilde{g}\mu_{B}B_{x}/v_{F}$,
the dependence of both $\tilde{g}$ and $v_{F}$ on the electron density
will directly influence the magnitude of the parallel field needed
to satisfy the nodal condition $\Delta kW=(2n+1)\pi/2$. Due to the
inverted nature of the bands, the g-factors in the conduction bands
are expected to evolve from $-20.5$ toward zero as the Fermi wavevector
decreases \cite{Konig2008}, while the expected magnitudes of the
Fermi velocities first decrease slightly and then more rapidly fall
to zero (Figure 4c, d). With these considerations we expect the magnitudes
of the induced pairing wavevectors at 1 T to fall to zero from values
near 1.2/$\mu$m as the Fermi wavevector decreases (Figure 4e). As
a result, the magnetic field needed to satisfy the nodal condition
increases as the electron density decreases, finally diverging at
zero electron density (calculated in black dotted lines in Figure
4a-b). Although the overall evolution agrees well with the expectation
from BHZ theory, we find that our measurements on niobium and aluminum
devices respectively yield values of $\tilde{g}/v_{F}$ that are approximately
1.9 and 1.4 times greater than those expected theoretically (see Supplement
VI).

Several aspects of the density-dependent data do not fall into the
modeling framework discussed above, and are interesting for further
consideration. First, we expect that the position of the node associated
with induced Cooper pair momentum should occur at higher parallel
magnetic field as the density is reduced, a behavior that we observe
only at high density. As the density is further reduced, the magnitude
of the nodal field eventually begins to decrease, an element of our
model that is not present and remains to be understood, but could
possibly be explained by a finite g-factor at zero density. Second,
in the aluminum device, the region of reduced resistance occuring
above the first node appears to be strongest near top gate voltages
equal to $-0.9$ V and $0.5$ V. We observe that these two regions
of reduced resistance are connected by a region in which the resistance
is more weakly reduced, but we have no reason to expect that the reduction
in resistance above the first node should vary as the density decreases.

Our measurements demonstrate that a parallel magnetic field can be
used both to tune the momentum of Cooper pairs in a material and to
clarify the nature of spin-orbit coupling in that material. A major
current goal of condensed matter physics is to understand the nature
of the superconductivity that results when electron pairing is combined
with materials possessing exotic spin textures. Therefore our new
understanding that the superconducting order parameter can be engineered
in space may be utilized to investigate spin physics within a broad
range of materials including InAs-based quantum wells or the surfaces
of three-dimensional topological insulators. Our method to tune the
Josephson energy could find additional application in the field of
quantum information processing, where direct control of the energy
levels in a single superconducting qubit could provide a powerful
tool for the investigation and optimization of qubit coherence.

\bibliographystyle{naturemag}
\bibliography{cpmomentumbib}

\begin{doublespace}
\textbf{Acknowledgments: }We acknowledge Ewelina M. Hankiewicz and
Grigory Tkachov for theoretical discussions. This work was supported
by the NSF DMR-1206016, by the STC Center for Integrated Quantum Materials
under NSF Grant No. DMR-1231319 and by Microsoft Corporation Project
Q. We acknowledge additional financial support from the German Research
Foundation (The Leibniz Program, Sonderforschungsbereich 1170 \textquotedblleft Tocotronics\textquotedblright{}
and Schwerpunktprogramm 1666), the EU ERC-AG program (Project 3-TOP)
and the Elitenetzwerk Bayern IDK \textquotedblleft Topologische Isolatoren\textquotedblright .

\textbf{Author Contributions: }The experiment is a collaboration between
the Harvard and Würzburg experimental groups. S.H., H.R., M.K., G.
B.-S., B. I. H., and A.Y. carried out the theoretical modeling and
analysis.

\textbf{Author Information: }The authors declare no competing financial
interests. Correspondence and requests for materials should be addressed
to yacoby@physics.harvard.edu.

\begin{figure}[p]
\centering{}\includegraphics[scale=0.35]{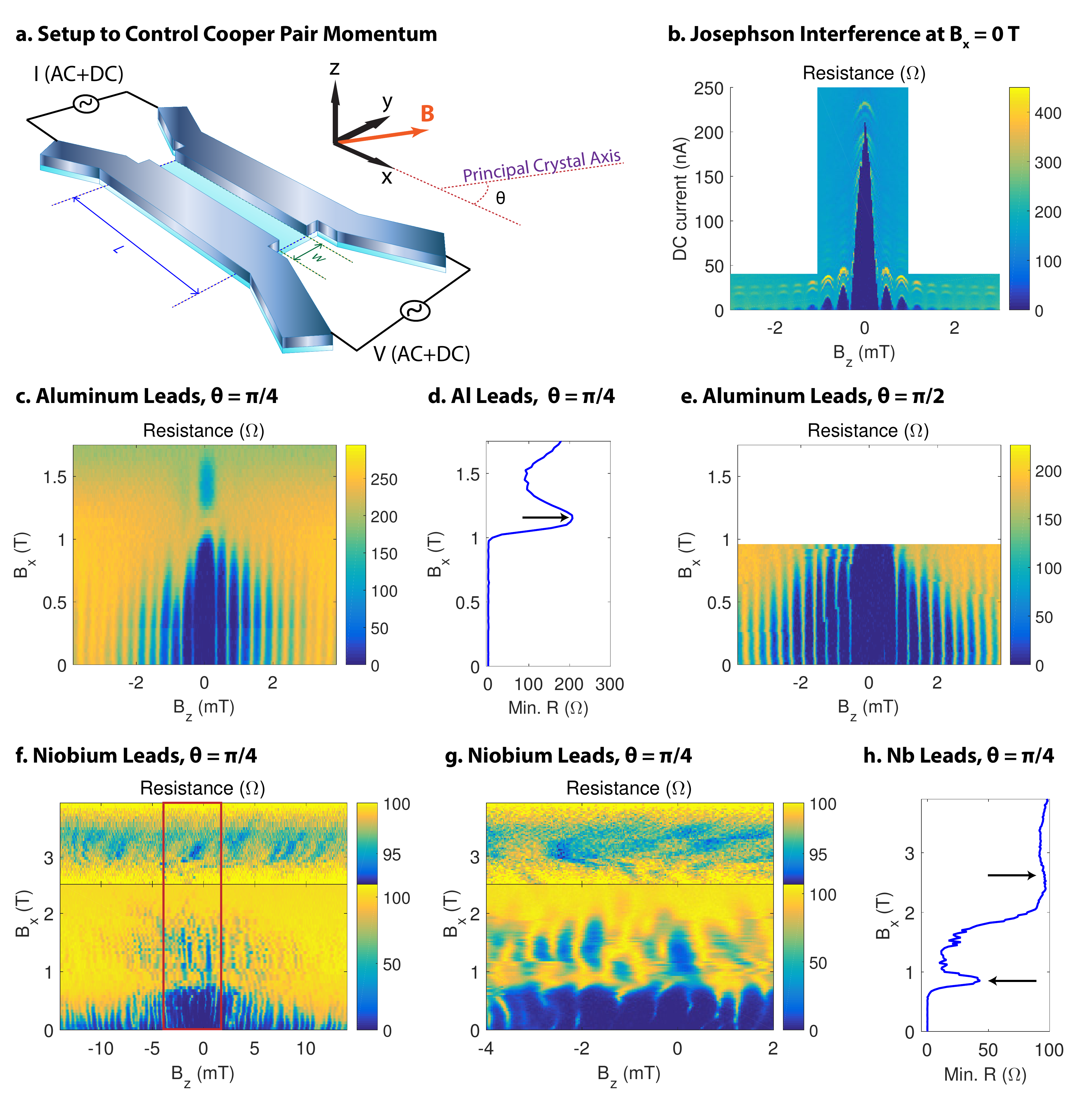}\caption{Experimental control of the order parameter and of pairing momentum.
\textbf{(a)} Two superconducting leads, composed of either aluminum
or niobium, couple to a rectangular section of HgTe quantum well to
form a Josephson junction. The width $W$ separating the leads is
always 800 nm, while the length $L$ of the junction is always 4 microns.
The resistance of the junction is monitored by applying a small AC
current bias (typically $\sim1$ nA) and concurrently measuring the
resulting AC voltage. Further sourcing DC current allows measurement
of critical currents and normal device resistance. The external magnetic
field $\vec{B}$ contains a small $z$ component to generate Josephson
interference, while here the larger $x$ component couples significantly
to the spin-degree of freedom. Junctions may be oriented at an angle
$\theta$ with respect to the $[100]$ principal crystal axis, modulo
$\pi/2$. \textbf{(b)} In the electron-doped regime, the devices show
Josephson interference consistent with transport through a doped bulk.
In all subsequent measurements, zero DC current is applied. \textbf{(c)}
The differential resistance of a junction with aluminum leads oscillates
due to Josephson interference as the perpendicular field varies. Increasing
the parallel field modulates the strength of induced superconductivity.
\textbf{(d)} Plotting the minimum resistance at each value of $B_{x}$
demonstrates the presence of a nodal resistance maximum near $B_{x}=1.1$
T. \textbf{(e)} In an aluminum-based device oriented with $\theta=\pi/2$,
increasing the parallel field similarly modifies the resistance. \textbf{(f)}
In a junction with niobium leads, a similar modulation of the resistance
occurs. \textbf{(g)} A more detailed study of the space outlined in
red in (f) highlights three regions of decreased resistance separated
by bands of high resistance near $B_{x}=0.9$ T and $B_{x}=2.7$ T.
In both (f) and (g), the decreased resistance above 2.5 T is highlighted
via a stretched color scale. \textbf{(h)} The minimum resistance at
each value of $B_{x}$ further shows the oscillatory nature of the
superconductivity as the parallel field increases. Successively higher
nodes (marked by arrows) occupy broader regions of parallel field,
while superconductivity also weakens as the parallel field increases. }
\end{figure}

\end{doublespace}

\begin{figure}[p]
\begin{centering}
\includegraphics[scale=0.4]{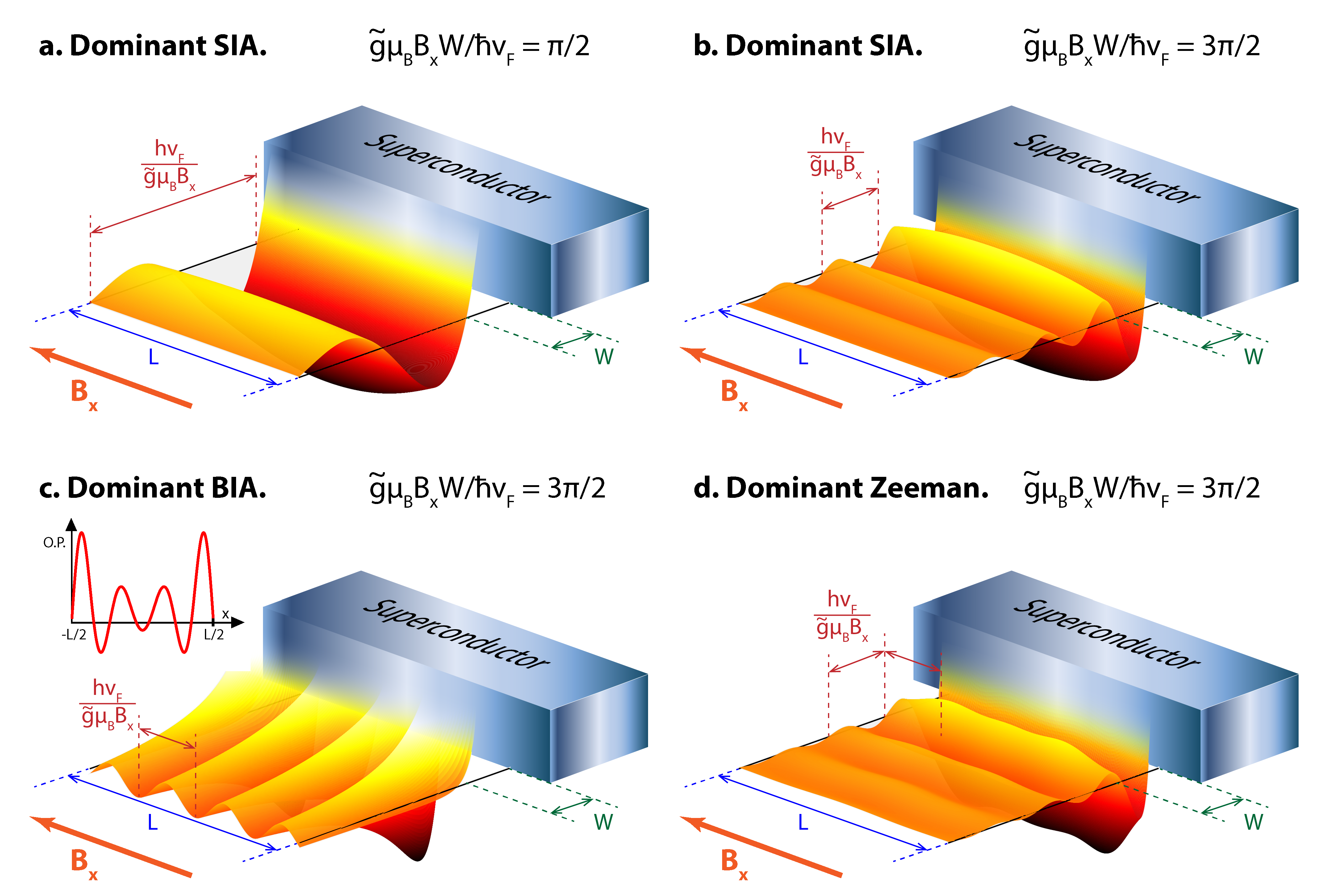}
\par\end{centering}

\caption{Theoretical prediction for the spatially varying order parameter near
a single superconducting lead, with $B_{z}=0$. \textbf{(a)} With
dominant SIA, application of an in-plane magnetic field $B_{x}$ induces
oscillations of the order parameter in the $y$ direction, with wavelength
$hv_{F}/\tilde{g}\mu_{B}B_{x}$. When $\tilde{g}\mu_{B}B_{x}W/\hbar v_{F}=\pi/2$
the first node of the order parameter occurs a distance $W$ from
the superconductor. \textbf{(b)} As the magnitude of magnetic field
increases, the wavelength of order parameter oscillations decreases.
When $\tilde{g}\mu_{B}B_{x}W/\hbar v_{F}=3\pi/2$, the second order
parameter node lies a distance $W$ from the superconductor. \textbf{(c)}
If instead BIA dominates, the order parameter oscillations occur in
the $x$ direction. As the magnetic field increases, the frequency
of oscillations increases. In the inset, a linecut of the order parameter
a distance $W$ from the superconductor demonstrates that oscillations
are an end effect, with amplitudes which decay into the bulk of the
2DEG. \textbf{(d)} With weak spin-orbit coupling, the parallel field
$B_{x}$ introduces order parameter oscillations in both directions.}
\end{figure}

\begin{figure}[p]
\begin{centering}
\includegraphics[scale=0.35]{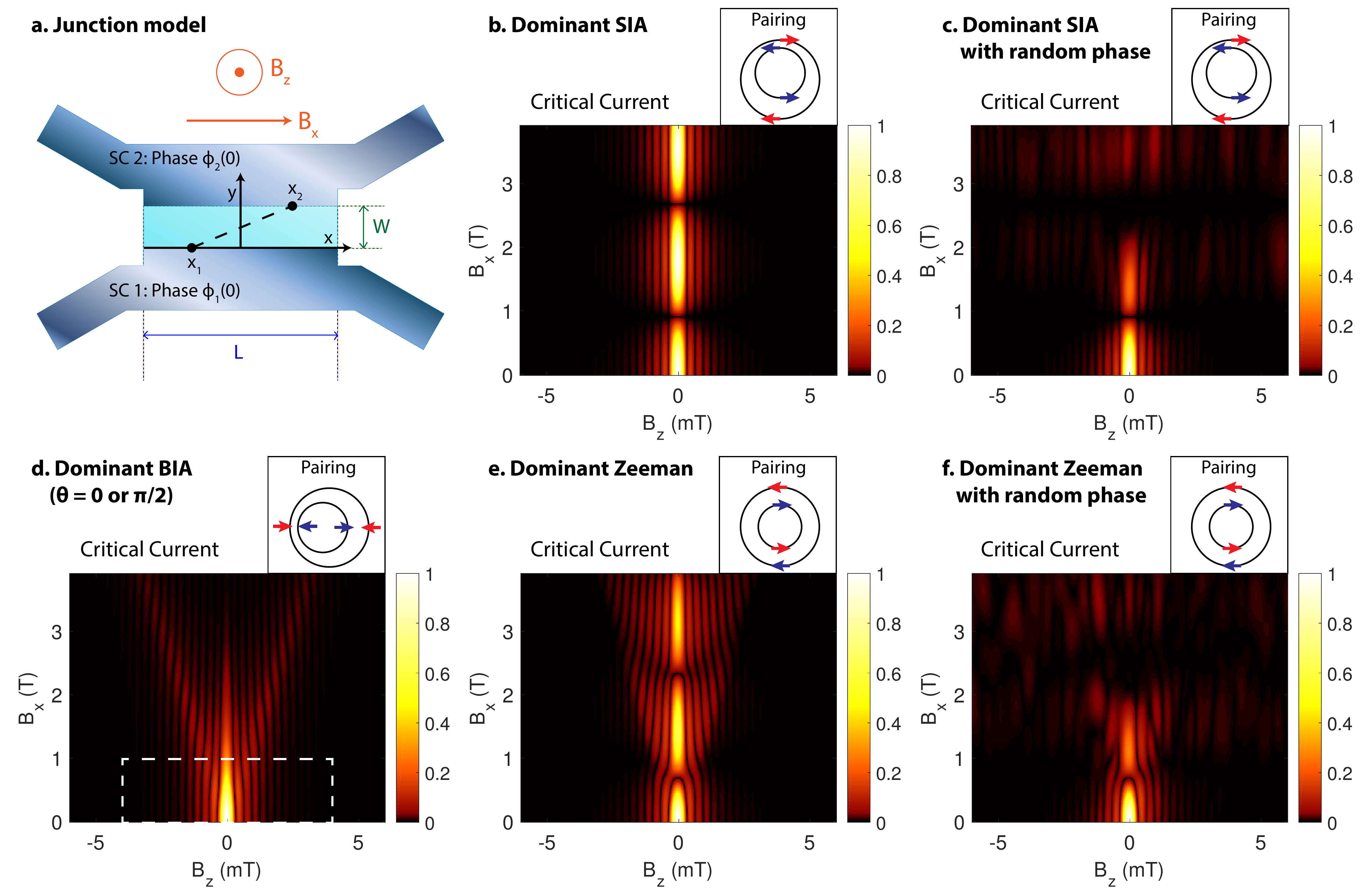}
\par\end{centering}

\caption{Modeling Josephson interference between two superconducting leads.
\textbf{(a)} With two leads, paired electrons may traverse the junction
beginning at a point $x_{1}$ in the lower superconducting lead (SC
1). The pairing amplitude at the point $x_{2}$ in the upper lead
(SC 2), takes account of the phase accumulated due to finite pairing
momentum within the HgTe quantum well. \textbf{(b)} With SIA dominant,
the external magnetic field $B_{x}$ increases the pairing wavevector
$\Delta k$ only in the $y$ direction. At certain values $\Delta k=(2n+1)\pi/(2W)$,
$n$ integer, the superconducting interference disappears. A diagram
schematically depicts the expected Fermi surfaces and Cooper pairing,
where arrows denote spin direction and pairs are each colored blue
or red. Similar diagrams throughout this figure indicate the expectation
for pairing and Fermi surfaces as the model parameters change. \textbf{(c)}
Randomness at the interface between the quantum well and superconductors
may arise due to structural imperfections. The random phase causes
superconductivity to weaken and increases the width of nodes as the
parallel field increases. \textbf{(d)} For junctions aligned to a
principal crystal axis, dominant BIA leads to a pairing wavevector
$\Delta k$ that grows in the $x$ direction as $B_{x}$ increases.
The critical current maxima then occur at increasingly large values
of $\left|B_{z}\right|$ as $B_{x}$ grows. Fabricating devices at
varying angles with respect to the crystal is expected to modify the
interference when BIA dominates. The region outlined in dashed white
corresponds to the measured region in Figure 1e. \textbf{(e)} With
dominant Zeeman coupling, the pairing magnitude is isotropic in-plane
and the interference grows as a hybrid of the SIA and BIA cases. Characteristically,
interference fringes repeatedly combine to form the central fringe
at each successive node in the parallel field. Additionally, with
zero perpendicular field, superconductivity disappears at values of
parallel field that are smaller than the nodal magnetic field in the
case with dominant SIA. \textbf{(f)} Including randomness leads to
a similar picture to (c), while retaining the combining of fringes
characteristic of dominant Zeeman coupling.}
\end{figure}

\begin{figure}[p]
\centering{}\includegraphics[scale=0.5]{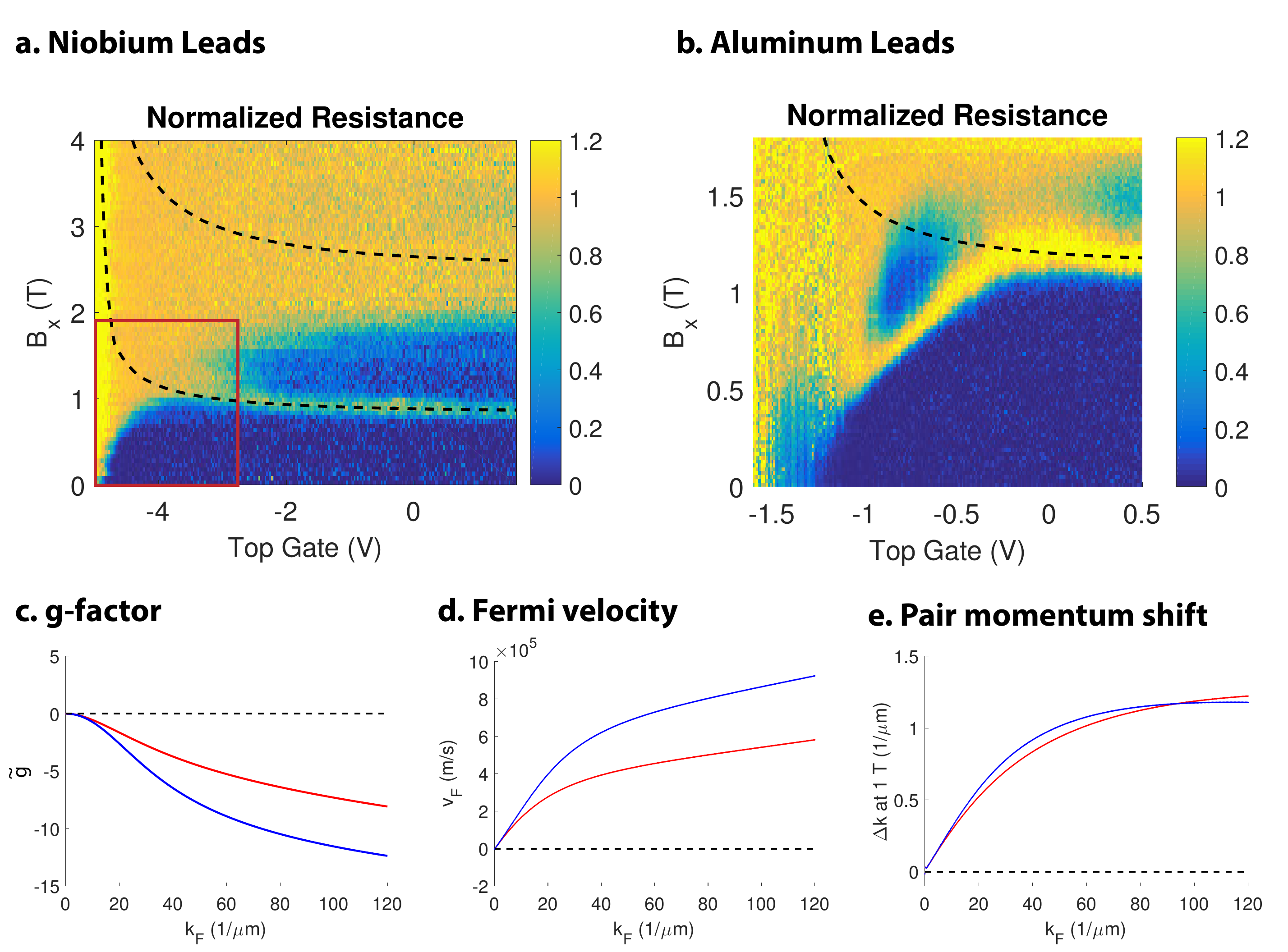}\caption{The evolution of minimum differential resistance as density and parallel
magnetic field $B_{x}$ vary. Differential resistance measurements
are normalized at each point by the normal junction resistance. \textbf{(a)}
At the highest gate voltage in the niobium junction, increasing the
magnetic field leads to periodic high-resistance nodes separating
regions of decreased resistance. As the gate voltage is decreased,
the magnetic field at the first node rises to larger values of $B_{x}$.
\textbf{(b)} The aluminum junction behaves similarly to the niobium
junction, although the measurement is limited to a smaller region
outlined in red in (a). At low density, the magnitude of the nodal
magnetic field begins to decrease as the density is lowered, a feature
which remains to be understood. \textbf{(c)} The values of $B_{x}$
at which we expect nodes to appear are sensitive to the density dependence
of both the in-plane g-factors and the Fermi velocities, calculated
here assuming that SIA is due to an electric field of 10 mV/nm. Blue
and red curves correspond to the inner and outer Fermi surfaces, respectively.
As the magnitude of the Fermi wavevector $k_{F}$ decreases, the in-plane
g-factors shift from -20.5 toward zero. \textbf{(d)} Meanwhile the
magnitudes of the Fermi velocities decrease to zero. \textbf{(e)}
The pairing momenta induced at 1 T consequently decrease to zero from
approximately 1.2/$\mu$m. Since there can be no induced momentum
at zero density, the nodal magnetic field diverges as the density
is lowered. The nodal magnetic field, averaged over the two Fermi
surfaces, is calculated using BHZ theory and plotted as dashed black
lines in (a) and (b).}
\end{figure}

\pagebreak{}

\begin{doublespace}
\noindent \begin{center}
{\LARGE{}Supplementary Information for Controlled Finite Momentum}{\large{}}\\
{\LARGE{} Pairing and Spatially Varying Order Parameter in}{\large{}}\\
{\LARGE{} Proximitized HgTe Quantum Wells}{\large{}}\\

\par\end{center}{\large \par}

\noindent \begin{center}
{\large{}Sean Hart$^{\dagger1}$, Hechen Ren$^{\dagger1}$, Michael
Kosowsky$^{1}$,}\\
{\large{} Gilad Ben-Shach$^{1}$, Philipp Leubner$^{2}$, Christoph
Brüne$^{2}$, Hartmut Buhmann$^{2}$, }\\
{\large{} Laurens W. Molenkamp$^{2}$, Bertrand I. Halperin$^{1}$,
Amir Yacoby$^{1}$}\\

\par\end{center}{\large \par}

\noindent \begin{center}
{\large{}$^{1}$Department of Physics, Harvard University, Cambridge,
MA, USA }\\
{\large{}$^{2}$ Physikalisches Institut (EP3), Universität Würzburg,
97074 Würzburg, Germany}\\
{\large{}$^{\dagger}$These authors contributed equally to this work}
\par\end{center}{\large \par}
\end{doublespace}

\section*{I. Wafer characteristics and General Measurements}

Junctions were fabricated using HgTe/HgCdTe heterostructures grown
in the $[001]$ crystal direction (the $z$ direction), composed as
shown in Supplementary Figure 1. Wafer I contained an 8 nm quantum
well with an electron density of $13.5\times10^{11}/\text{cm}^{2}$
and a mobility of $390,000\text{ cm}^{2}/\text{Vs}$. Wafer II contained
a 7.8 nm quantum well with an electron density of $2.9\times10^{11}/\text{cm}^{2}$
and a mobility of $790,000\text{ cm}^{2}/\text{Vs}$. Josephson junctions
fabricated on these wafers were aligned at varying angles with respect
to the $[110]$ and $[1\bar{1}0]$ cleavage edges of the crystal,
but we do not know which is which in our samples. Therefore, we can
only specify that the angular alignment corresponds to a rotation
angle $\theta$ with respect to the $[100]$ crystal axis, modulo
$\pi/2$. Although we do not know which principal axis $\theta$ is
referenced to experimentally, our model predicts the same results
when $\theta$ is referenced to either. The $x$ and $y$ axes always
lie respectively perpendicular and parallel to the direction of current
flow in devices (see Figure 1 of main text). 

Throughout this supplement, devices are referred to in the following
manner. Device A was fabricated by depositing aluminum leads onto
a mesa etched into Wafer I, and was oriented at an angle $\theta=\pi/4$.
Devices B, C, and D were concurrently fabricated by depositing aluminum
leads on Wafer II, and were respectively oriented at different angles
$\theta=0,\pi/2,$ and $\pi/4$. Device E was fabricated using wafer
I, contained niobium leads, and was oriented at $\theta=\pi/4$. Device
F was fabricated using Wafer I, contained aluminum leads, and was
oriented at $\theta=\pi/4$. Note that data from devices C, D and
E were reported in the main text.

Devices were processed as follows. To define the ends of the junctions,
mesas 100 nm in height were etched using an argon ion source. To fabricate
superconducting contacts, the contact area was etched enough to expose
the quantum well using argon milling. Without breaking vacuum, the
contact material was then deposited. For devices A-D and F, 5 nm of
titanium was deposited by thermal evaporation, followed by thermal
evaporation of 15 nm of aluminum. For device E, 10 nm of titanium
was deposited by e-beam evaporation, followed by 130 nm of niobium
deposited by DC magnetron sputtering. Next, a 50 nm layer of aluminum
oxide was grown using atomic layer deposition, to isolate the final
topgate layer (10 nm of titanium and 250 nm of gold) from the underlying
junctions.

Measurements were performed primarily in a dilution refrigerator outfitted
with a 6-1-1 vector magnet capable of applying up to 6 T in one direction,
and 1 T in the two remaining directions. Experiments were carried
out at the mixing chamber base temperature of 50 mK. Unless otherwise
mentioned, all measurements occurred in this system. One measurement
was performed in a separate dilution refrigerator with a base temperature
of 10 mK.

As either the external magnetic field or the topgate voltage was varied,
the differential four-terminal resistance of junctions was monitored
using standard lock-in techniques. To determine critical currents
or to provide bias sufficient to measure the normal resistance of
the junctions, sometimes a DC current bias was applied to the junctions.
Otherwise all measurements occurred with no DC current bias.

\section*{II. Characterization of Thin Aluminum Leads}

The resistance of thin aluminum leads was characterized as a function
of magnetic field in both the $x$ direction and the $y$ direction
(Supplementary Figure 2). In each plot, red and blue color coding
corresponds to the two leads of a single junction. Different junctions
were used in the two plots, demonstrating the consistency of the fabrication
process. The critical parallel field of the 15 nm aluminum films was
consistently above 1.5 T, in agreement with a previous study of thin
aluminum \cite{Meservey1971}. The data presented in Supplementary
Figure 2b was collected at a temperature of 10 mK in the system discussed
previously.

\section*{III. Critical Current of a Josephson Junction Under External Parallel
Magnetic Field}

Here we present measurements of the critical current in device D,
for different values of the external magnetic field. When compared
to corresponding measurements of differential resistance without any
applied DC current (Supplementary Figure 4e and Supplementary Figure
5e), it is apparent that both measurement modes reveal the same basic
behaviors.

In the measurements of critical current presented here, increasing
the external field in either the $x$ or $y$ direction results in
a decrease of the maximum critical current (Supplementary Figure 3a,
b). This decrease occurs more rapidly for $B_{y}$, where the critical
current becomes too small to reliably measure when $B_{y}$ exceeds
0.44 T. $B_{x}$, however, must exceed 1.1 T before critical currents
become immeasurably small.

Upon normalization of the interference pattern at each separate value
of parallel field, the asymmetry between the two directions becomes
more pronounced (Supplementary Figure 3c, d). At all values of $B_{x}$,
the shape of the Josephson interference remains essentially unaffected.
By contrast, as $B_{y}$ increases the critical current splits into
two separate maxima which occur at larger values of $\left|B_{z}\right|$.
In Supplement XI we model the effect of the in-plane field $B_{y}$,
which we expect to induce a finite $x$ component of the pairing wavevector
that grows linearly with $B_{y}$. In junctions with finite length
and with $B_{z}=0$, this pairing momentum in the $x$ direction leads
to oscillations in the order parameter which are most pronounced near
the ends of the junction. As a consequence, we expect that as $B_{y}$
increases, the maximum critical currents in our junctions will occur
at values of $\left|B_{z}\right|$ that grow linearly with $B_{y}$.
With only the parallel field $B_{x}$ present, however, the induced
pair momentum is expected to lie along the $y$ direction. In this
case no such wavevector in the $x$ direction is observed.

\section*{IV. Josephson Junctions Rotated with Respect to the Crystal}

Measurements of the differential resistance were performed on junctions
oriented at different angles with respect to the crystal lattice,
in order to determine whether bulk inversion asymmetry (BIA) plays
a significant role in the momentum acquired by Cooper pairs. As previously
mentioned, these devices A-E have orientation $\theta=\pi/4,0,\pi/2,\pi/4,$
and $\pi/4$ respectively. Devices A-D use aluminum leads, while device
E uses niobium leads. For each set of the devices, we explored behaviors
resulting from a parallel magnetic field applied in the $x$ direction
or in the $y$ direction, which in the previous section were shown
for device D to differ.

Even as the angle $\theta$ varies among devices, the manner in which
superconductivity evolves due to the applied field $B_{x}$ remains
qualitatively unchanged at high density (Supplementary Figure 4).
Devices B-D, fabricated on a single piece of wafer, show quantitative
agreement in the value of $B_{x}$ at which a superconducting node
appears. Devices A and E were separately fabricated on Wafer I, and
show slight quantitative differences but nevertheless the same shape.
The appearance of these nodes in the interference evolution, with
no dependence on the crystal orientation, signals that structural
inversion asymmetry (SIA) dominates the behavior of our quantum wells
in the electron-doped regime (Supplement VI-IX).

With the parallel field applied in the $y$ direction and at high
density, the interference pattern splits, forming a `V' shape as $B_{y}$
increases that is qualitatively identical for all values of $\theta$
(Supplementary Figure 5). The slope of the two arms of the `V' varies
among devices, but is similar for devices B-D which were fabricated
concurrently. The most dramatic effect is seen for device E with 130
nm thick niobium leads, in which the slope is approximately 7 times
smaller than the other devices.

From the above measurements one can conclude that the basic differences
in interference as $B_{x}$ or $B_{y}$ is increased have little to
do with the orientation of the crystal lattice. The most striking
difference is found among the data with the magnetic field oriented
along the $y$ direction, in which the thickness of the leads correlates
to the slope of the interference splitting. This behavior, which results
from magnetic flux penetrating the area $dL$ formed by the length
$L$ of the junction and the height difference $d$ between the center
of the quantum well and the center of the leads, is modeled in Supplement
XI.

\section*{V. Evolution of Interference Lobes as the Parallel Magnetic Field
$B_{x}$ Increases, in a Device with $\theta=\pi/2$}

In the previous section, it was observed that at particular values
of the parallel magnetic field $B_{x}$, nodes of suppressed superconductivity
occur in our Josephson junctions. Additionally, this evolution of
the Josephson interference with the parallel field $B_{x}$ does not
depend on the device orientation with respect to the crystal. The
appearance of nodes with lack of $\theta$ dependence already suggests
that the effect of bulk inversion asymmetry (BIA) in our devices is
small, and that structural inversion asymmetry (SIA) dominates (Supplement
VI-IX). Still, the evolution of critical currents, in particular the
maximum critical current of interference lobes occurring at nonzero
perpendicular field, can provide further evidence that BIA is weak.
When $\theta=\pi/2$, in the limit of strong BIA these side lobe maximum
critical currents are expected to grow as the parallel field $B_{x}$
is increased from zero (Figure 3d of main text). However, if SIA is
strong, these critical currents are expected to monotonously decay
as the parallel field increases up to the first node.

In Device C, oriented at $\theta=\pi/2$, we study the evolution of
the first three side lobes adjacent to the central lobe (Supplementary
Figure 6a). A measurement of the critical currents of these lobes
indicates that all lobes are largest when $B_{x}=0$ T (Supplementary
Figure 6b). We extract the maximum critical current for each side
lobe, plotted in Supplementary Figure 6c for each lobe. All critical
currents are largest when $B_{x}=0$ T, and all decay monotonously
until becoming indistinguishable from zero. This evolution of the
side lobe critical currents provides additional evidence consistent
with weak BIA in our devices.

\section*{VI. Four-Band Model and Spin-Orbit Effects in the Quantum Well}

A four-band model has been developed starting from $k\cdot p$ theory,
and subsequently used to describe the topology of HgTe quantum wells
\cite{Bernevig2006_Science}. Here we adopt an elaboration on this
model, where bulk inversion asymmetry (BIA), structural inversion
asymmetry (SIA), and coupling to an external magnetic field are included.
The four bands originate in the $s-$ and $p-$like bands of the underlying
crystals, so that the basis states are written as $\left|E1,m_{J}=+1/2\right\rangle $,
$\left|H1,m_{J}=+3/2\right\rangle $, $\left|E1,m_{J}=-1/2\right\rangle $,
and $\left|H1,m_{J}=-3/2\right\rangle $. In this notation, $E1$
refers to electron-like states with angular momentum 1/2, while $H1$
refers to hole-like states with angular momentum 3/2. The Hamiltonian
describing the system is then \cite{Rothe2010,Weithofer2013,Konig2008}:

\begin{equation}
\begin{array}{cc}
H_{1}= & \epsilon\left(\hat{k}^{2}\right)+M\left(\hat{k}^{2}\right)s_{z}+A\hat{k}_{x}s_{x}\sigma_{z}-A\hat{k}_{y}s_{y}+h\left(\cos(2\theta^{th})\sigma_{y}+\sin(2\theta^{th})\sigma_{x}\right)s_{y}\\
 & +R_{0}\frac{s_{z}+1}{2}\left(\hat{k}_{x}\sigma_{y}-\hat{k}_{y}\sigma_{x}\right)-\mu_{B}\cdot\vec{B}\cdot\vec{M}
\end{array}
\end{equation}

where

\begin{equation}
\epsilon\left(\hat{k}^{2}\right)=C-D\left(\hat{k}_{x}^{2}+\hat{k}_{y}^{2}\right),M\left(\hat{k}^{2}\right)=M-B\left(\hat{k}_{x}^{2}+\hat{k}_{y}^{2}\right)
\end{equation}

and

\begin{equation}
M_{x}=\left(\begin{array}{cc}
 & g_{||}\\
g_{||}
\end{array}\right),M_{y}=\left(\begin{array}{cc}
 & -ig_{||}\\
ig_{||}
\end{array}\right),M_{z}=\left(\begin{array}{cccc}
g_{E\perp}\\
 & g_{H\perp}\\
 &  & -g_{E\perp}\\
 &  &  & -g_{H\perp}
\end{array}\right).
\end{equation}

The first four terms are those present in the original theory of Bernevig,
Hughes, and Zhang. The fifth term describes the magnitude $h$ of
the BIA and the angle $\theta^{th}$ between the $x$ axis and the
$[100]$ crystal direction. The sixth term includes SIA with strength
$R_{0}/(e\mathcal{E})$, where $\mathcal{E}$ is the magnitude of
an electric field oriented along the $z$ axis. Coupling to the external
magnetic field occurs anisotropically and is different for $E1$ and
$H1$ states due to the inability of magnetic field to couple $m_{J}=\pm3/2$
to first order. The value of the parallel g-factor $g_{||}$ and perpendicular
g-factors $g_{E\perp}$ and $g_{H\perp}$, along with all other parameter
values, are listed in Table 1. The in-plane wavevectors are $\hat{k}_{x}=i\partial_{x}$
and $\hat{k}_{y}=i\partial_{y}$. 

The bare effect of a parallel magnetic field is visualized by setting
$h=0$ and $R_{0}=0$ and calculating the conduction band spectrum,
under various values of $\vec{B}$ (Supplementary Figure 7a-c). In
each plot, the bands are plotted up to the Fermi energy, which varies
in order to demonstrate various limiting cases of weak or strong spin-orbit
coupling. The Fermi surfaces are projected onto a plane below the
bands, with the spin texture plotted at various points on the Fermi
surfaces. The $x$ and $y$ components of the spin vectors are equal
respectively to $\left\langle \sigma_{x}\left(1+s_{z}\right)/2\right\rangle $
and $\left\langle \sigma_{y}\left(1+s_{z}\right)/2\right\rangle $,
the spin expectation values projected onto the $E1$ bands. 

With no external magnetic field, the conduction band is doubly degenerate
due to the presence of the spin degree of freedom, modeled here up
to a Fermi energy equal to 20 meV. Changing the magnetic field to
3 T in the $x$ direction lifts this degeneracy due to the Zeeman
effect, resulting in two concentric Fermi surfaces centered at $\vec{k}=0$.
Within the outer Fermi surface, spins point toward $-x$, due to the
negative value of $g_{||}$. Within the inner Fermi surface, spins
are oppositely oriented. Application of a 3 T magnetic field in the
$y$ direction identically splits the bands, although the spins in
this case orient along $\pm y$. It can be seen that with no spin-orbit
coupling and an applied in-plane magnetic field, pairs may form with
nonzero wavevector in any in-plane direction.

Including dominant SIA (equivalent to a perpendicular electric field
of $10$ mV/nm) with no BIA modifies the expectation for the band
structure (Supplementary Figure 7d-f). Even without any external magnetic
field, in this case the bands are spin-split for all nonzero wavevectors.
The axial symmetry of the spin-orbit coupling results in the spin
texture shown in Supplementary Figure 7e, modeled up to a higher Fermi
energy of 50 meV to highlight the limit of strong SIA. Now coupled
with an external magnetic field, these factors result in a behavior
that is different from the case with no spin-orbit coupling. When
magnetic field is applied in the $x$ ($y$) direction, the two Fermi
surfaces shift oppositely in the $y$ ($x$) direction. Consistent
with our observations on aluminum-basd junctions in the electron-doped
regime, this shift implies that a magnetic field in the $x$ direction
introduces a pairing wavevector in the $y$ direction. Also superficially
consistent with our data, a magnetic field in the $y$ direction leads
to a pairing wavevector in the $x$ direction. However, the magnitude
of this wavevector alone is too small to explain the experimental
evolution of interference as $B_{y}$ is increased. A full explanation
of this effect is presented in Supplementary Information XI. Finally,
the lack of dependence of SIA on the angle $\theta^{th}$ is consistent
with our measurements (Supplementary Figures 4 and 5).

In the remaining case, the effect of dominant BIA with no SIA is investigated
in Supplementary Figure 7g-i, for an angle $\theta^{th}=0$ and up
to a Fermi energy of 20 meV. Similarly to the case of strong SIA,
with no external magnetic field present the bands are spin-split at
nonzero wavevectors. However, the texture of spins at the Fermi surfaces
displays tetrahedral symmetry in this case. As a result, when parallel
magnetic field is present in the $x$ ($y$) direction, the two Fermi
surfaces shift oppositely in the $x$ ($y$) direction. This shifting
is orthogonal to the shifts present with strong SIA, and does not
agree with our interference measurements on devices aligned with $\theta=0$
or $\pi/2$. Furthermore, the shift direction rotates by $\pi/2$
as the angle $\theta^{th}$ becomes $\pi/4$. This prediction that
the direction of induced Cooper pair momentum should depend on $\theta^{th}$
is also inconsistent with our results.

Finally, in the main text, we model the evolution with density of
the in-plane g-factor $\tilde{g}$, the Fermi velocity $v_{F}$, and
the pair momentum shift $\Delta k\approx\tilde{g}\mu_{B}B_{x}/\hbar v_{F}$,
under the assumption that BIA is absent and SIA is dominant (equivalent
to a perpendicular electric field of 10 mV/nm). At each density, a
particular value of the magnetic field satisfies the condition $\Delta kW=\pi/2$,
leading to a node in the induced superconductivity in the junction.
We find that the evolution with density of this nodal magnetic field
value agrees with our model at high densities. In niobium and aluminum
devices respectively, the value of the nodal magnetic field is consistent
with values of of $\tilde{g}/v_{F}$ that are approximately 1.9 and
1.4 times greater than those expected theoretically. Using different
values for the electric field only weakly modifies this conclusion.
For example, if the SIA was instead equivalent to a perpendicular
electric field of 40 mV/nm, we would find values of of $\tilde{g}/v_{F}$
approximately 2.1 and 1.5 times greater than the theoretical expectation,
only a $\sim10\%$ difference.

\section*{VII. Model of a Two-Dimensional Electron Gas Contacted by Superconducting
Leads}

In the following sections, we model the coupling of superconducting
leads to our quantum well. We consider a geometry in which a two-dimensional
electron gas (2DEG) is contacted by a pair of superconducting leads
with a controlled phase difference between them, and we seek to calculate
the maximum supercurrent that can be carried between the strips. The
following is a more complete derivation of the pair propagator $F$
used in the main text to carry out this goal. We assume a Hamiltonian
$H=H_{0}+H_{2}$, where $H_{0}$ is the Hamiltonian for the 2DEG in
the absence of the superconductor, and $H_{2}$ is the coupling between
the superconductors and the 2DEG, described by a pairing Hamiltonian
of the form

\begin{equation}
H_{2}=-\int dxdy\left[\Delta(x,y)\Psi^{\dagger}(x,y)+\Delta^{*}(x,y)\Psi(x,y)\right].
\end{equation}

\noindent Here $\Psi(x,y)\equiv\psi_{\uparrow}(x,y)\psi_{\downarrow}(x,y)$
is an operator which annihilates a singlet pair of electrons in the
2DEG at the point $(x,y)$, while the pair potential $\Delta(x,y)$
is a complex number that depends on the phase of the superconductor
and the tunneling amplitude at that point.

We assume that the contacts between the 2DEG and the superconductors
occur at the edges of the superconductors, located at $y=0$ and $y=W$,
so that we may write

\begin{equation}
\Delta(x,y)=\lambda_{1}(x)\delta(y)+\lambda_{2}(x)\delta(y-W)\equiv\Delta_{1}(x,y)+\Delta_{2}(x,y),
\end{equation}

\noindent with $-L/2<x<L/2$. We assume that the magnitude of the
coupling is constant along each lead, but the phase will vary if there
is a perpendicular magnetic field $B_{z}\neq0.$ We choose a gauge
where the vector potential points in the $x$ direction, with $A_{x}=-B_{z}(y-W/2)$,
so that the vector potential vanishes along the midline of the 2DEG.
If the superconducting strips have identical widths $W_{SC}$, then
the couplings $\lambda_{j}$ will have the form

\begin{equation}
\lambda_{j}(x)=\left|\lambda_{j}\right|e^{2\pi i\phi_{j}(x)},
\end{equation}

\begin{equation}
\phi_{j}(x)=\phi_{j}(0)+\frac{(-1)^{j-1}xB_{z}(W+W_{sc})}{2\Phi_{0}},
\end{equation}

\noindent with $j=1,2$. The phases in equation (15) are determined
by the condition that there should be no net current flow along the
length of the superconducting strips, so the phase gradient in each
superconductor should be canceled by the vector potential along the
center line of the superconductor.

We assume that $B_{z}$ is sufficiently weak that the cyclotron radius
$R_{c}=\hbar k_{F}/eB_{z}$ is large compared to $W$ (typically in
our devices $R_{c}\approx10$ $\mu$m, which is large compared to
$W=800$ nm). In this case, we may ignore the effect of $B_{z}$ on
the trajectories of electrons in the 2DEG. Moreover, since we have
chosen the vector potential to vanish along the midline of the 2DEG,
an electron crossing from $y=0$ to $y=W$ will acquire no net phase
due to the vector potential. We also ignore, for the moment, any orbital
effects of the parallel field $B_{||}$. Thus the 2DEG Hamiltonian
$H_{0}$ will include the Zeeman coupling to $B_{||}$, as well as
the spin-orbit coupling, but will not include terms due to the magnetic
field in the kinetic energy.

To lowest order in the pairing Hamiltonian $H_{2}$, the portion of
the total energy that depends on the phase difference between the
two superconducting leads can be written in the form:

\begin{equation}
E=-\int dx_{2}\left[\lambda_{2}^{*}(x_{2})\left\langle \Psi(x_{2},W)\right\rangle _{1}+c.c.\right],
\end{equation}

\noindent where $\left\langle \Psi(x,y)\right\rangle _{1}$ is the
order parameter at point $(x,y)$ induced by the superconductor $j=1$.
In turn, this may be written in the form

\begin{equation}
\left\langle \Psi(x,y)\right\rangle _{1}=\int dx_{1}\lambda_{1}(x_{1})F(x,x_{1},y),
\end{equation}

\noindent where $F$ is the propagator from point $(x_{1},0)$ to
point $(x,y)$ for an induced Cooper pair. We will determine the form
of $F$ in the following section.

As a first approximation, we may ignore the fact that there are boundaries
of the 2DEG at $x=\pm L/2$ and that electrons will be reflected at
these boundaries (either specularly or diffusely, and possibly with
a spin flip). Similarly, we ignore the possibility of single-particle
reflection at $y=0$ or $y=W$, where the superconducting leads touch
the 2DEG. Furthermore, we assume that the electron density is constant
in the 2DEG, and we ignore any interactions between electrons in the
2DEG. We also ignore scattering by impurities inside the 2DEG. Then
the propagator $F$ may be calculated for an infinite, translationally
invariant 2DEG, where the momentum of each electron is a good quantum
number. We believe that corrections due to reflections at the boundaries
will have quantitative effects but will not affect qualitatively the
form of our results. Modeling of critical current including specular
reflections at mesa edges will be discussed in Supplement X and XI.

\section*{VIII. Derivation of a General Formula for the Pair Propagator $F$}

The four-band model Hamiltonian $H_{1}$ for our quantum wells was
previously discussed. Since we presently are interested in behavior
near the Fermi energy $E_{F}$, and within the conduction bands only,
in the following we adopt a simplified two-band model Hamiltonian.
This Hamiltonian for an electron in the 2DEG with momentum $\vec{k}$
is given by the $2\times2$ matrix

\begin{equation}
H_{\vec{k}}^{0}=v_{F}(k-k_{F})+\vec{\beta}(\hat{k})\cdot\vec{\sigma},
\end{equation}

\begin{equation}
\beta_{j}=\tilde{g}\mu_{B}B_{j}/2+\hat{k}_{i}S_{ij},
\end{equation}

\noindent where $B_{j}$ are the $x$ and $y$ components of the in-plane
magnetic field, $\tilde{g}$ is an effective g-factor, $\hat{k}\equiv\vec{k}/k$,
and $\hat{k}_{i}S_{ij}\sigma_{j}$ is the spin-orbit coupling term,
which we assume to be small compared to the Fermi energy. We have
here assumed a single electron band, and assumed that band structure
is isotropic in the absence of spin-orbit coupling. We can then write

\begin{equation}
H_{\vec{k}}^{0}=\sum_{\eta}\epsilon_{\vec{k}\eta}P^{\hat{k}\eta},
\end{equation}

\noindent with $\eta=\pm1$. Here $\epsilon_{\vec{k}\eta}$ are the
two eigenvalues, and $P^{\hat{k}\eta}$ are projection matrices given
by

\begin{equation}
\epsilon_{\vec{k}\eta}=v_{F}(k-k_{F})+\eta\left|\vec{\beta}\right|,P^{\hat{k}\eta}=(1+\eta\hat{\beta}\cdot\vec{\sigma})/2,
\end{equation}

\noindent with $\hat{\beta}\equiv\vec{\beta}/\left|\vec{\beta}\right|$.

We next define a $2\times2$ matrix function

\begin{equation}
g(\vec{r},\epsilon)\equiv\frac{1}{(2\pi)^{2}}\sum_{\eta}\int d^{2}ke^{i\vec{k}\cdot\vec{r}}\delta(\epsilon-\epsilon_{\vec{k}\eta})P^{\hat{k}\eta}.
\end{equation}

\noindent Then, letting $\vec{r}=(x-x_{1},y)$, the pair propagator
$F(x,x_{1},y)$ may be expressed as 

\begin{equation}
F(\vec{r})=\int_{0}^{\infty}d\epsilon\int_{0}^{\infty}d\epsilon'\frac{\text{\text{tr}}\left[g(\vec{r},\epsilon)\sigma^{y}g^{T}(\vec{r},\epsilon')\sigma^{y}+g(\vec{r},-\epsilon)\sigma^{y}g^{T}(\vec{r},-\epsilon')\sigma^{y}\right]}{2(\epsilon+\epsilon')},
\end{equation}

\noindent where $T$ indicates the matrix transpose.

We are interested in the situation where $k_{F}r\gg1$, and $\left|\epsilon\right|\ll E_{F}$.
Then the integration over the direction of $\vec{k}$ is dominated
by regions close to the end points where $\vec{k}$ is either parallel
or antiparallel to $\vec{r}$, and the expression for $g(\vec{r},\epsilon)$
may be approximated by

\begin{equation}
g(\vec{r},\epsilon)\approx\frac{k_{F}^{1/2}}{(2\pi)^{3/2}v_{F}r^{1/2}}\sum_{\eta}\left[e^{ik_{F}r}e^{i\epsilon r/v_{F}}e^{-i\pi/4}e^{-i\delta k_{\eta}^{+}r}P^{\eta+}+e^{-ik_{F}r}e^{-i\epsilon r/v_{F}}e^{i\pi/4}e^{i\delta k_{\eta}^{-}r}P^{\eta-}\right]
\end{equation}

\noindent where $P^{\eta\pm}$ is equal to $P^{\hat{k}\eta}$, with
$\hat{k}=\pm\hat{r}\equiv\pm\vec{r}/r$ and

\begin{equation}
\delta k_{\eta}^{\pm}=\frac{\eta\left|\vec{\beta}(\pm\hat{r})\right|}{v_{F}}.
\end{equation}

When we substitute the expression for $g$ in formula (23) for $F(\vec{r})$,
we may ignore the terms proportional to $e^{\pm2ik_{F}r}$, as these
rapidly oscillating terms will give vanishing contribution to the
energy if the width of the contacts between the 2DEG and the superconducting
strips are large compared to $1/k_{F}$. Performing the integrals
over $\epsilon$ and $\epsilon'$ in the remaining terms, one obtains
the result

\begin{equation}
F(\vec{r})\approx\frac{C}{r^{2}}\sum_{\eta\eta'}N_{\eta\eta'}(\hat{r})e^{-i\left(\delta k_{\eta}^{+}-\delta k_{\eta'}^{-}\right)r},
\end{equation}

\noindent with $C=k_{F}/(8\pi^{2}v_{F})$, and

\begin{equation}
N_{\eta\eta'}(\hat{r})=N_{\eta\eta'}(-\hat{r})=\frac{1-\eta\eta'\hat{\beta}(\hat{r})\cdot\hat{\beta}(-\hat{r})}{2}.
\end{equation}

\section*{IX. Special Cases and Limiting Forms of the Pair Propagator}

Here we discuss several special cases which lead to limiting forms
for the pair propagator $F(\vec{r})$. The above expressions (equations
(26) and (27)) may be simplified in the limit where the Zeeman energy
is small compared to the spin-orbit energy splitting. When $B_{||}=0$,
we find that $\hat{\beta}(\hat{r})=-\hat{\beta}(-\hat{r})$, so that
$N_{\eta\eta'}=\delta_{\eta\eta'}$. Furthermore, when $\eta=\eta'$,
we see that the exponent in equation (26) is equal to zero, so $F$
will have no oscillations as a function of $r$. If $B_{||}$ is nonzero
but still small compared to the spin-orbit splitting, it remains a
good approximation to set $N_{\eta\eta'}=\delta_{\eta\eta'}$. In
the exponent, however, we have

\begin{equation}
\left(\delta k_{\eta}^{+}-\delta k_{\eta}^{-}\right)=\eta\tilde{g}\mu_{B}\vec{B}_{||}\cdot\vec{\beta}(\vec{r}).
\end{equation}

An important example is the case of pure SIA spin-orbit coupling,
where the matrix $S$ has the Rashba form, $S\propto i\tau^{y}$,
where $\tau^{y}$ is a Pauli matrix. In this case we may write

\begin{equation}
\left(\delta k_{\eta}^{+}-\delta k_{\eta}^{-}\right)r=\eta\Delta\vec{k}\cdot\vec{r}
\end{equation}

\noindent with

\begin{equation}
\Delta\vec{k}=\hat{z}\times\vec{B}_{||}\tilde{g}\mu_{B}/v_{F}.
\end{equation}

The oscillations in $F(\vec{r})$ have a simple interpretation in
this case. When $B=0$, the Fermi surface consists of two circles
centered about the origin, split by the spin-orbit coupling, with
spin orientations shown in Figure 3b of the main text. Application
of a weak in-plane magnetic field will shift the two Fermi circles
in opposite directions, by amounts $\pm\Delta\vec{k}/2$. The function
$F(\vec{r})$ describes the propagator when a singlet pair of electrons
is injected at one point and removed at a second point, separated
by $\vec{r}$. For large separations, $F(\vec{r})$ is dominated by
pairs of electrons that are close to the Fermi energy, with wave vectors
opposite to each other and parallel or antiparallel to $\vec{r}$.
Because the two electrons must have opposite spins, they must belong
to the same branch of the Fermi surface. Thus, the induced pairs will
have total momenta equal to $\pm\Delta\vec{k}$, depending on the
branch $\eta$. The momentum shifts are manifest in the phase factors
$e^{i\eta\Delta\vec{k}\cdot\vec{r}}$, which appear in $F(\vec{r})$
in this case.

In the case of pure BIA coupling, the matrix $S$ is $\propto\tau^{z}$,
in our coordinate system. We may again write the phase accumulation
in the form (29), but now the direction of $\Delta\vec{k}$ depends
on the directions of $\vec{B}_{||}$ relative to the crystal axes.

The formula for $F(\vec{r})$ also becomes simple in the case where
the Zeeman energy is large compared to the spin-orbit splitting. In
this case, the Fermi surface consists of two concentric circles, with
spin that are uniformly aligned on each circle, either parallel or
antiparallel to $\vec{B}_{||}$. In order to form a spin singlet,
we must choose one electron from each Fermi circle. If we also require
that the momenta be parallel or antiparallel to $\vec{r}$, we see
that the induced electron pair will have a total momentum equal to
$\pm\hat{r}\tilde{g}\mu_{B}B_{||}/v_{F}.$ Thus we should find that
the phase shift is independent of the direction of $\vec{r}$.

These expectations may be confirmed using the formulas derived above.
In the case where the Zeeman energy is large compared to the spin-orbit
splitting, we find that $\vec{\beta}(\hat{k})$ is independent of
$\hat{k}$, and thus $N_{\eta\eta'}=\delta_{\eta,-\eta'}$. Furthermore,
$\delta k_{\eta}^{+}-\delta k_{-\eta}^{-}=\eta\tilde{g}\mu_{B}B_{||}/v_{F},$
independent of $\vec{r}$.

\section*{X. Reflections from the Sample Edges}

Taking into account the effects of electron reflections from the ends
of the sample, at $x=\pm L/2$, we should rewrite the propagator $F$
in a more general form as

\begin{equation}
F(x_{2},x_{1},W)=F_{0}(\vec{r})+F_{1}\left[\left(x_{2}+\frac{L}{2}\right),\left(x_{1}+\frac{L}{2}\right),W\right]+F_{2}\left[\left(\frac{L}{2}-x_{2}\right),\left(\frac{L}{2}-x_{1}\right),W\right],
\end{equation}

\noindent where $F_{0}$ is the function given by equation (26) for
the infinite system, $\vec{r}=(x_{2}-x_{1},W),$ as before, and $F_{1}$
and $F_{2}$ describe the contributions of electrons reflected from
the left boundary or right boundary respectively. We assume that the
length $L$ is long enough that we can neglect the effects of electrons
that scatter multiple times from opposite boundaries. Here we will
assume that the boundaries at $x=\pm L/2$ are represented by infinite
potential barriers, which are perfectly smooth, so that electrons
are specularly reflected with no change in spin. The symmetry of our
problem will then be such that $F_{1}$ and $F_{2}$ have identical
functional forms, so we need only find the form of $F_{1}$. For convenience,
we move the left boundary to the line $x=0$, and we assume that the
right boundary is located at $x=\infty$. Using similar reasoning
to what we used in the translationally invariant case, we may write
$F_{1}$ in the form

\begin{equation}
F_{1}(x_{2},x_{1},W)=\int_{0}^{\infty}d\epsilon\int_{0}^{\infty}d\epsilon'\frac{\text{tr}\left[h(x_{2},x_{1},W,\epsilon)\sigma^{y}h^{T}(x_{2},x_{1},W,\epsilon')\sigma^{y}+c.c.\right]}{2(\epsilon+\epsilon')}
\end{equation}

\noindent where

\begin{equation}
\begin{aligned}h(x_{2},x_{1},W,\epsilon) & =-\frac{k_{F}^{1/2}}{(2\pi)^{3/2}v_{F}s^{1/2}}\sum_{\eta_{1},\eta_{2}}[e^{ik_{F}s}e^{i\epsilon s/v_{F}}e^{-i\pi/4}e^{-i(\delta k_{1,\eta_{1}}^{+}s_{1}+\delta k_{2,\eta_{2}}^{+}s_{2})}P_{1}^{\eta_{1}+}P_{2}^{\eta_{2}+}\\
 & +e^{-ik_{F}s}e^{-i\epsilon s/v_{F}}e^{i\pi/4}e^{i(\delta k_{1,\eta_{1}}^{-}s_{1}+\delta k_{2,\eta_{2}}^{-}s_{2})}P_{1}^{\eta_{1}-}P_{2}^{\eta_{2}-}],
\end{aligned}
\end{equation}

\noindent where $s=[(x_{1}+x_{2})^{2}+W^{2}]^{1/2}$, $s_{1}=sx_{1}/(x_{1}+x_{2})$,
$s_{2}=s-s_{1}$, and 

\begin{equation}
\delta k_{j,\eta}^{\pm}=\frac{\eta\left|\vec{\beta}(\pm\hat{k}_{j})\right|}{v_{F}},
\end{equation}

\noindent for $j=1,2,$ with

\begin{equation}
\hat{k}_{1}=-\frac{[(x_{1}+x_{2}),-W]}{s},\hat{k}_{2}=\frac{[(x_{1}+x_{2}),W]}{s}.
\end{equation}

\noindent Furthermore, we have 

\begin{equation}
P_{j}^{\eta\pm}=(1+\eta\hat{\beta}(\pm\hat{k}_{j})\cdot\vec{\sigma})/2.
\end{equation}

We now turn to one particular example. In order to evaluate expression
(32) for $F_{1}$, we must first evaluate the trace over a product
of projection matrices and $\sigma^{y}$. In the case of strong SIA
coupling and weak magnetic field, the trace simplifies, and we obtain
the result

\begin{equation}
\text{tr}\left[P_{1}^{\eta_{1}+}P_{2}^{\eta_{2}+}\sigma^{y}\left(P_{1}^{\eta_{3}-}P_{2}^{\eta_{4}-}\right)^{T}\sigma^{y}\right]=\delta_{\eta_{1}\eta_{3}}\delta_{\eta_{2}\eta_{4}}\left[\sin^{2}\theta\delta_{\eta_{1}\eta_{2}}+\cos^{2}\theta\delta_{\eta_{1},-\eta_{2}}\right],
\end{equation}

\noindent where $\sin\theta=W/s$. Furthermore for $B_{||}$ in the
$y$ direction, we find

\begin{equation}
(\delta k_{j,\eta}^{+}-\delta k_{j,\eta}^{-})s_{j}=(-1)^{j}\eta x_{j}\frac{\tilde{g}\mu_{B}B_{y}}{v_{F}}.
\end{equation}

\noindent Thus, in the case of strong SIA and $B_{||}$ in the $y$
direction, we find

\begin{equation}
F_{1}(x_{2},x_{1},W)=\frac{2C\left[\sin^{2}\theta\cos\Delta k(x_{1}-x_{2})+\cos^{2}\theta\cos\Delta k(x_{1}+x_{2})\right]}{(x_{1}+x_{2})^{2}+W^{2}},
\end{equation}

\noindent where $\Delta k=\tilde{g}\mu_{B}B_{x}/v_{F},$ and the constant
$C$ is the same as in equation (26).

\section*{XI. Modeling Josephson Interference}

Using the pair propagator $F(\vec{r})$ and equation (16), we can
calculate the Josephson energy and critical current for our junctions.
In the limit of either strong BIA or strong SIA, the Cooper pair momentum
shift occurs at an angle $\alpha$ with respect to the $x$ axis and
the pair propagator is

\begin{equation}
F(x_{2},x_{1},W)=\frac{k_{F}}{8\pi{}^{2}v_{F}}\cdot\frac{e^{i\gamma}+e^{-i\gamma}}{(x_{2}-x_{1})^{2}+W^{2}},\gamma=\Delta k\left(\sin(\alpha)W+\cos(\alpha)(x_{2}-x_{1})\right).
\end{equation}

\noindent As previously noted, in this case pairing occurs internally
to each Fermi surface. In the limit of weak spin-orbit coupling, the
pair propagator instead takes the form

\begin{equation}
F(x_{2},x_{1},W)=\frac{k_{F}}{8\pi{}^{2}v_{F}}\cdot\frac{e^{i\gamma}+e^{-i\gamma}}{(x_{2}-x_{1})^{2}+W^{2}},\gamma=\Delta k\sqrt{W^{2}+(x_{2}-x_{1})^{2}}.
\end{equation}

\noindent Due to the opposite spin polarization of the two Fermi surfaces,
pairing in this limit is expected to occur between Fermi surfaces,
in contrast to the limit of large spin-orbit coupling.

The Josephson energy $E$ is obtained in each limit by evaluating
equation (16). By differentiating the Josephson energy with respect
to the phase difference $\phi_{1}(0)-\phi_{2}(0)$ we find the current-phase
relation of the junction, which is then maximized with respect to
the phase difference to obtain the critical current.

In the main text we consider only a parallel field along the $x$
direction, and we model the effect of structural imperfections in
the interface between the superconducting leads and the quantum well.
These would introduce random fluctuations in the direction of the
in-plane magnetic field at each interface, leading to a random component
of the pairing momentum in the $x$ direction that grows linearly
with the in-plane field. Hence, we introduce a random phase $\chi\propto\left(R_{1}(x_{1})-R_{2}(x_{2})\right)B_{x}$,
where the random variables $R_{1}(x_{1})$ and $R_{2}(x_{2})$ correspond
to fluctuations in the direction of the parallel field at each interface.
The modeled step size in $x$ is 40 nm, with no correlations between
adjacent positions. The random phase $\chi$ is uniformly distributed
between zero and an upper bound whose absolute magnitude is equal
to 15\% of the maximum phase generated by the intrinsic momentum.
With this randomness, the calculated critical currents diminish in
magnitude as the in-plane field increases, and nodes occur over a
larger range of in-plane field, in agreement with the experimental
observation (shown in Figure 3c of the main text for the case of dominant
SIA).

In general, the parallel field $\vec{B}_{||}$ can be oriented anywhere
in the plane, which modifies $\alpha$ accordingly in the case that
spin-orbit coupling is strong. Additionally, loosening the constraint
that $\vec{B}_{||}$ lie parallel to $x$ introduces an artifact wavevector
$q_{y}\approx2\pi B_{||}\sin(\beta)d/\Phi_{0}$, where $d$ is the
height difference between the centers of the quantum well and of the
superconducting leads, and $\beta$ is the angle between the parallel
magnetic field and the $x$ axis. This additional phase arises due
to the magnetic flux penetrating the area $dL$ formed between the
leads and the quantum well due to this height difference. Importantly,
no flux penetrates this area when the parallel component of magnetic
field is only in the $x$ direction, so that in this case the pair
momentum is solely determined by the Zeeman coupling and the spin-orbit
coupling.

The behavior of Josephson interference in our devices essentially
involves different mechanisms when the parallel magnetic field lies
in the $x$ or $y$ direction. With the above modeling it is clear
that this difference is due to the dependence of $q_{y}$ on the magnetic
field angle $\beta$, so that the data for the magnetic field $B_{y}$
reflects primarily the difference in height between the superconducting
leads and the quantum well. Since we cannot entirely rule out either
dominant Zeeman coupling or dominant SIA, we model both possibilities
for this field direction. With dominant SIA and the height difference
$d$ set to either 10 nm (Supplementary Figure 8a) or 70 nm (Supplementary
Figure 8b), the model agrees well with the experimental observation.
The corresponding model without any spin-orbit coupling also agrees,
however (Supplementary Figure 8c, d).

Nevertheless, it is clear that with the parallel magnetic field in
the $y$ direction, the most prominent feature in the response of
the device is driven by the parallel magnetic flux penetrating the
area $dL$ and not by effects intrinsic to the heterostructure. Assuming
that only this parallel magnetic flux contributes, one can estimate
the distance $d$ for each device, accounting for the slight difference
in $W_{SC}$ for aluminum and niobium devices (1 $\mu$m and 400 nm
respectively). For devices A-E, the corresponding distance $d\approx21,10,9,7,$
and $70$ nm, in agreement with lithographic dimensions. The similar
values of $d$ for devices B-D reflects the fact that these devices
were all fabricated concurrently. Device E, in which niobium was used
for the leads, has a much larger value of $d$ due to the fact that
the niobium thickness was larger than the aluminum thicknesses.

Athough the parallel magnetic flux dominates the response of devices
to the field $B_{y}$, with purely SIA it is still in principle possible
in this direction to extract the intrinsic nature of spin-orbit coupling.
Since the wavevectors $q_{y}$ and $\Delta k$ add and subtract, the
`V' shape of supercurrent evolution contains two nearly identical
slopes, which in our measurements are unobservable due to the concurrent
decay of superconductivity. However, normalizing the theoretical critical
current magnitude still reveals the possibility to determine the nature
of spin-orbit coupling using this parallel field direction (Supplementary
Figure 8e).

An additional characteristic common among the data sets is an asymmetry
in the interference pattern upon inversion of one component of the
applied magnetic field. In Supplementary Figure 8f we show an interference
pattern measured on Device F with both positive and negative components
$B_{y}$ and $B_{z}$. Here the data appears invariant under inversion
of both components of the magnetic field, as we expect from time-reversal
symmetry. However, the lack of symmetry under inversion of a single
component of the magnetic field suggests that devices lack structural
symmetry under rotation by 180 degrees. We may model this asymmetry
as arising from a difference in lengths of the two leads on either
side of the junction. An exaggeration of this effect, where the interface
to one lead is 4 microns and to the other is 4.5 microns, shows increased
intensity of interference for positive perpendicular field as compared
to negative perpendicular field (Supplementary Figure 8g), similar
to what is observed experimentally.

Finally, in all of the above modeling we have ignored contributions
due to reflections at the ends of the mesa. In Supplementary Figure
8h, we plot the expected evolution of interference upon increasing
$B_{y}$ assuming dominant SIA and with $d=0$, ignoring the possibility
of reflections at mesa boundaries. We may include specular reflections
at the mesa boundaries in the model, as discussed in Supplement X.
With these contributions, the interference evolution is quantitatively
modified (Supplementary Figure 8i). However, the `V' shape of the
interference evolution is still present, with each arm of the `V'
having the same slope as was obtained by ignoring edge reflections.
Hence, we conclude that the contribution of specular reflections preserving
the spin direction only quantitatively modifies the expected device
behavior. We have not carried out calculations for other boundary
conditions, such as diffuse reflection, but we expect that results
in these cases would not be qualitatively different from the cases
of specular reflection or no reflection at all.

\begin{table}[p]
\noindent \begin{centering}
\begin{tabular}[b]{ccccccccccc}
{\footnotesize{}$A$ (eV$\cdot$Å)} & {\footnotesize{}$B$ (eV$\cdot$Å$^{2}$)} & {\footnotesize{}$D$ (eV$\cdot$Å$^{2}$)} & {\footnotesize{}$M$ (eV)} & {\footnotesize{}$h$ (eV)} & {\footnotesize{}$\theta^{th}$} & {\footnotesize{}$R_{0}/(e\mathcal{E})$ (nm$^{2}$)} & {\footnotesize{}$\mathcal{E}$ (mV/nm)} & {\footnotesize{}$g_{E\perp}$} & {\footnotesize{}$g_{H\perp}$} & {\footnotesize{}$g_{||}$}\tabularnewline
\hline 
{\footnotesize{}3.645} & {\footnotesize{}-68.6} & {\footnotesize{}-51.2} & {\footnotesize{}-0.010} & {\footnotesize{}0, 0.0016} & 0 & {\footnotesize{}-15.6} & {\footnotesize{}0, 10} & {\footnotesize{}22.7} & {\footnotesize{}-1.21} & {\footnotesize{}-20.5}\tabularnewline
\end{tabular}
\par\end{centering}

\caption{List of parameters used to model the band structure. Parameters correspond
to a quantum well in the inverted regime, with a well width of 70
Å.}
\end{table}

\pagebreak{}

\date{\noindent 
\begin{figure}[p]
\begin{centering}
\includegraphics[scale=0.6]{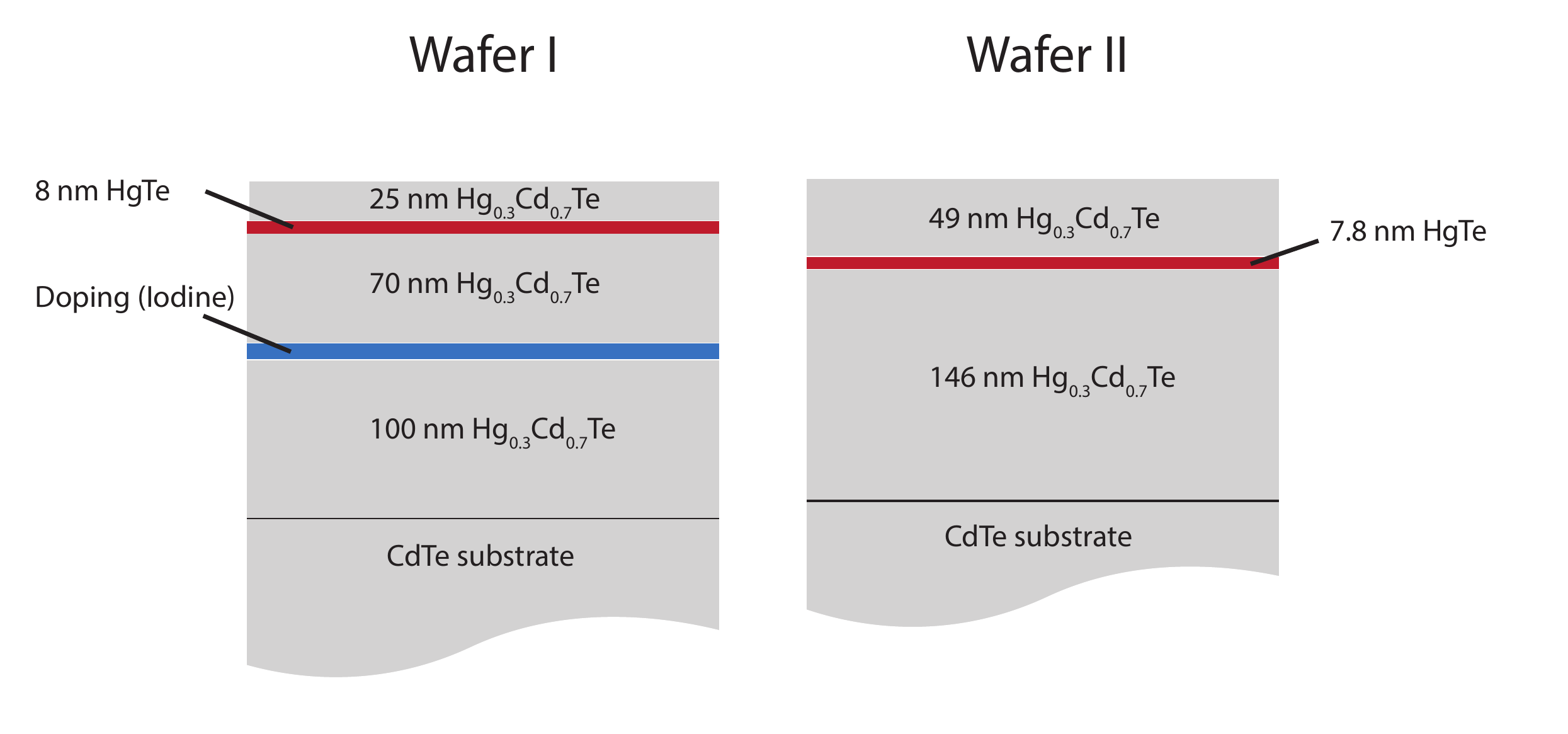}
\par\end{centering}

Supplementary Figure 1: Composition of the heterostructures used in
the experiment, labeled Wafer I and Wafer II. Both wafers consist
of a HgTe quantum well surrounded by barriers of Hg$_{0.3}$Cd$_{0.7}$Te.
In Wafer I, a layer containing iodine dopants lies 70 nm below the
quantum well. Both wafers were grown on CdTe substrates, in the $[001]$
crystal direction.
\end{figure}
}

\date{\noindent 
\begin{figure}[p]
\begin{centering}
\includegraphics[scale=0.5]{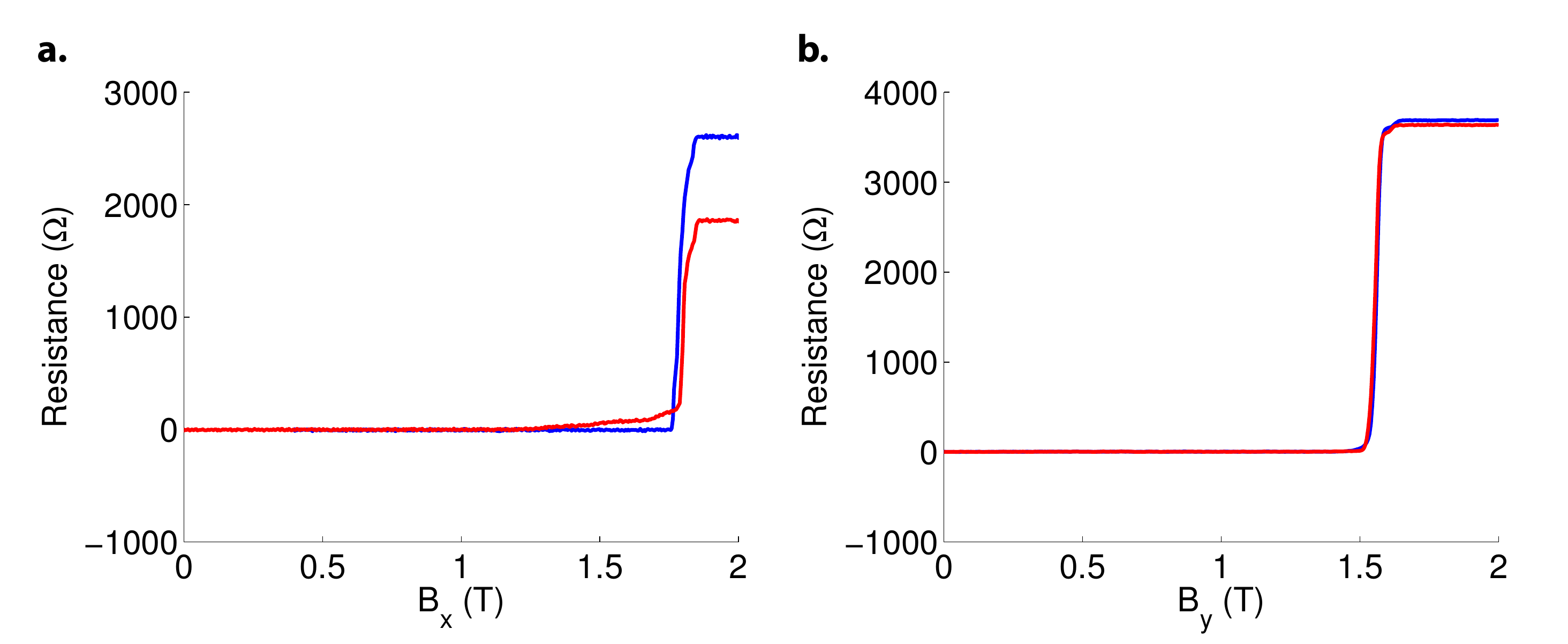}
\par\end{centering}

Supplementary Figure 2: Resistances of junction leads as the parallel
magnetic field is increased in the $x$ and $y$ directions, for junctions
with aluminum leads. The critical field in the plane of the leads
is consistently above 1.5 T.
\end{figure}
}

\date{\noindent 
\begin{figure}[p]
\begin{centering}
\includegraphics[scale=0.5]{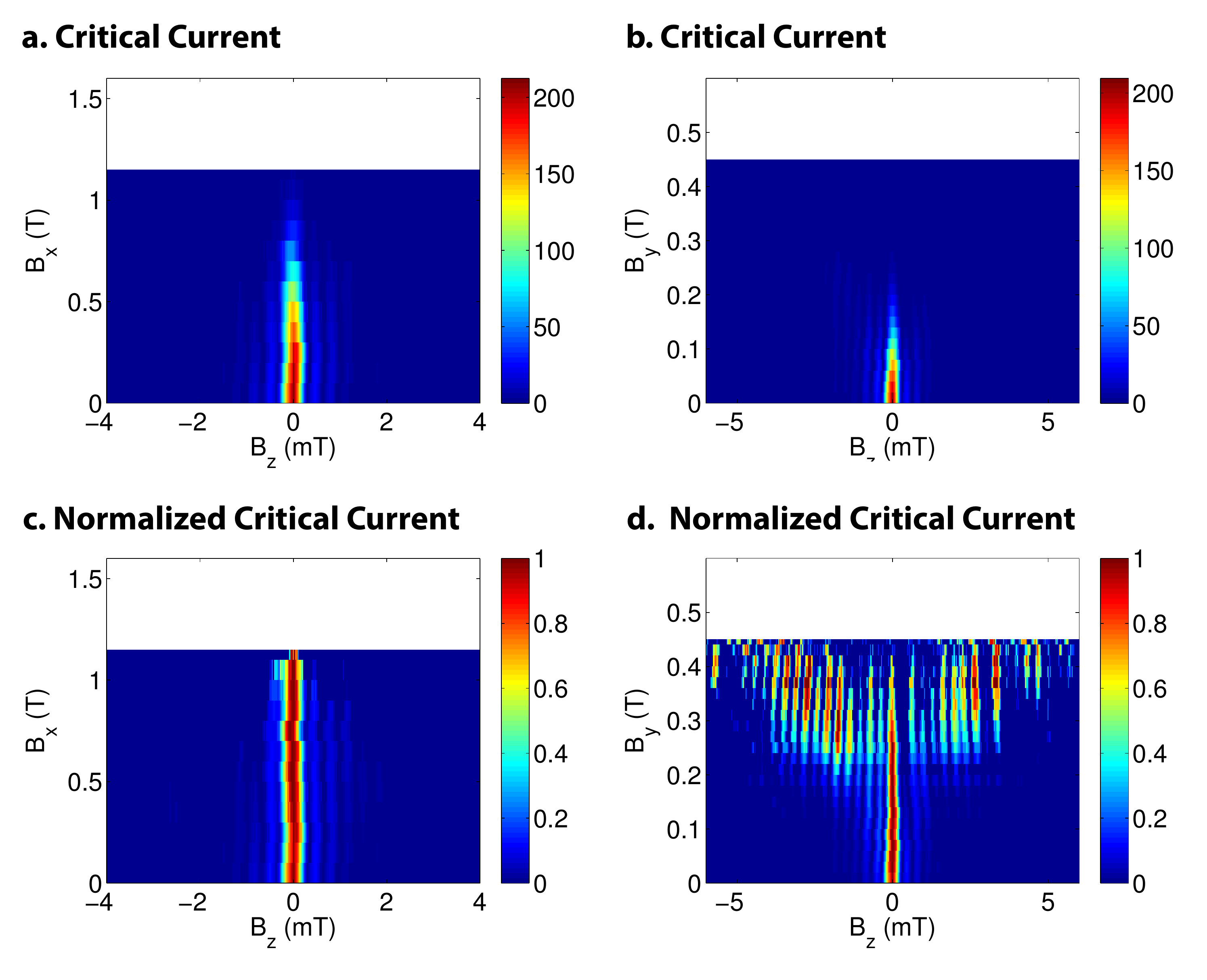}
\par\end{centering}

Supplementary Figure 3: The critical current as a function of perpendicular
magnetic field $B_{z}$, as parallel components of the magnetic field
are varied. The data presented here was taken using the device with
aluminum leads presented in the main text (device D). \textbf{a, b)}
As the parallel magnetic field in either the $x$ or $y$ direction
is increased, the magnitude of the maximum critical current decreases.
This decrease occurs more rapidly in $y$ direction than in the $x$
direction. \textbf{c)} Normalizing the Fraunhofer interference at
each value of $B_{x}$ shows that the shape of the interference pattern
remains essentially unaffected until it becomes immeasurably small.
\textbf{d)} In the $B_{y}$ direction, normalization reveals a dramatically
different behavior of the Fraunhofer interference, where critical
current maxima occur at higher values of $B_{z}$ as $B_{y}$ is increased.
Concurrently, the weight of the critical current at $B_{z}=0$ mT
decreases to 0. These observations match those deduced through measurements
of the differential resistance as the parallel magnetic field varies
in either the $x$ or $y$ direction. Therefore both differential
resistance and critical current measurements reflect the same basic
phenomenon.
\end{figure}
}

\date{\noindent 
\begin{figure}[p]
\begin{centering}
\includegraphics[scale=0.4]{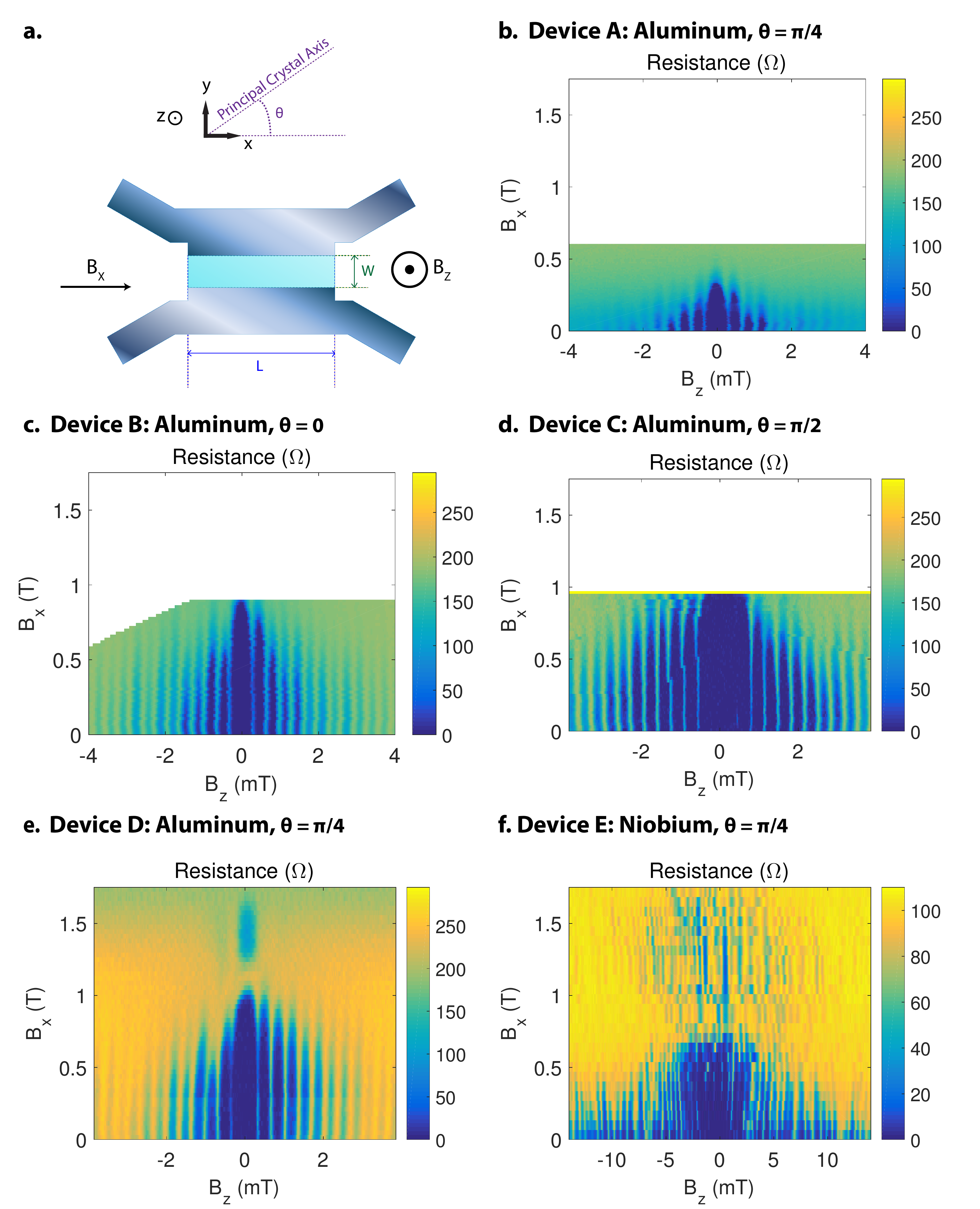}
\par\end{centering}

Supplementary Figure 4: Josephson interference as the magnetic field
$B_{x}$ is increased. \textbf{a)} Junctions were oriented at an angle
$\theta$, modulo $\pi/2$, with respect to the $[100]$ axis of the
crystal. Devices generating data in (c-e) were concurrently fabricated
with respect to the same crystal cleavage edge. Devices in (b) and
(f) were separately fabricated. All junctions had aluminum leads except
for in (f), where niobium leads were used. \textbf{b)} For a junction
oriented at $\theta=\pi/4$ with respect to the crystal, the differential
resistance is monitored as both the perpendicular field $B_{z}$ and
the parallel field $B_{x}$ are altered. As $B_{x}$ increases, the
position of nodes in the interference pattern does not change, but
the interference gradually disappears. \textbf{c)} A junction oriented
at $\theta=0$ and \textbf{d)} a junction oriented at $\theta=\pi/2$
with respect to the crystal show qualitatively identical behavior.
\textbf{e)} A further junction aligned at $\theta=\pi/4$ shows the
same behavior, as previously presented in the main text. \textbf{f)}
Also in the main text, the junction with niobium leads is oriented
with $\theta=\pi/4$ and shows interference which remains strongly
weighted at $B_{z}=0$ T. The observations on aluminum devices are
all consistent with dominant SIA in the quantum well.
\end{figure}
\begin{figure}[p]
\begin{centering}
\includegraphics[scale=0.4]{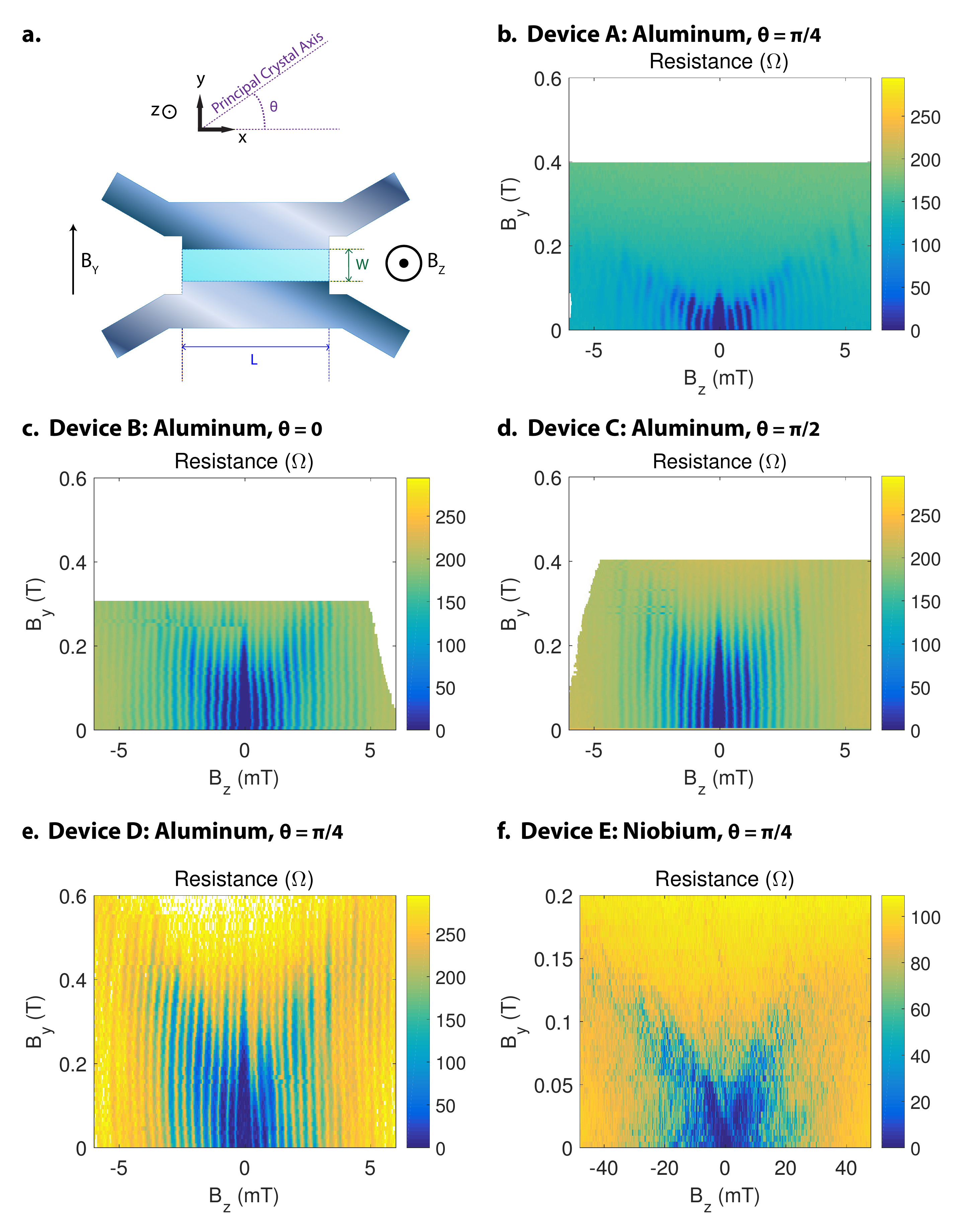}
\par\end{centering}

Supplementary Figure 5: Josephson interference as the magnetic field
$B_{y}$ is increased. \textbf{a)} Junctions were oriented at an angle
$\theta$, modulo $\pi/2$, with respect to the $[100]$ axis of the
crystal. As in Supplementary Figure 4, devices generating data in
(c-e) were concurrently fabricated with respect to the same crystal
cleavage edge, and devices in (b) and (f) were separately fabricated.
\textbf{b)} With the junction aligned such that $\theta=\pi/4$, the
differential resistance is monitored as a function of the perpendicular
field $B_{z}$ and the parallel field $B_{y}$. Increasing $B_{y}$
rapidly causes the weight of interference fringes to shift to larger
$B_{z}$ values, forming a `V' shape. The interference evolves more
rapidly due to a parallel field in the $y$ direction than in the
$x$ direction due to the fact that leads are spatially displaced
in $z$ with respect to the quantum well. \textbf{c)} Orienting a
junction at $\theta=0$ introduces no qualitative change to the behavior,
as is also the case with \textbf{d)} a junction oriented at $\theta=\pi/2$.
Junctions from the main text with \textbf{e)} aluminum and \textbf{f)}
niobium leads are presented, also displaying similar behavior. The
enhanced scale of $B_{z}$ in the niobium-based junction is due to
the increased thickness of the leads.
\end{figure}
}

\date{\noindent 
\begin{figure}[p]
\begin{centering}
\includegraphics[scale=0.5]{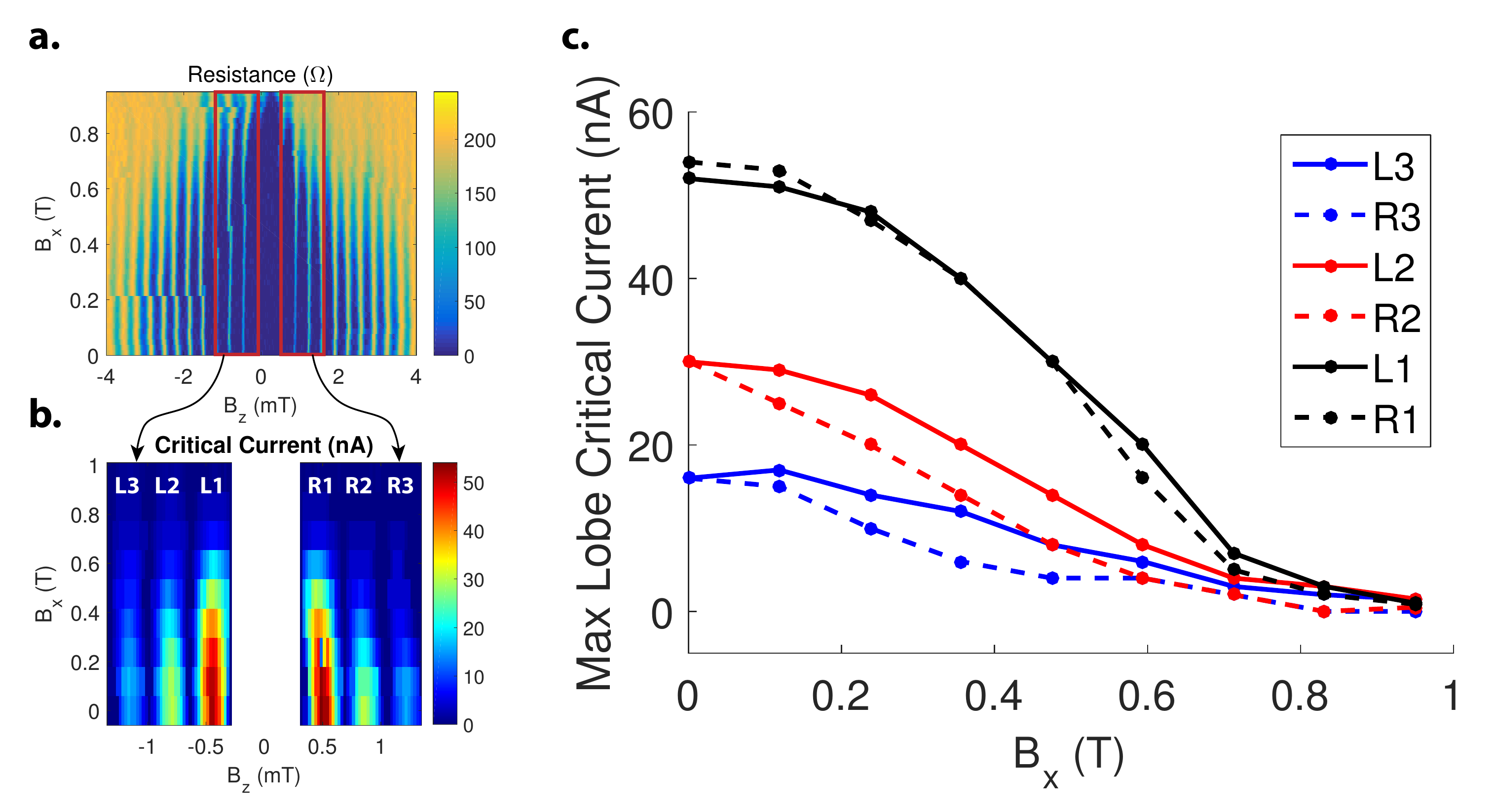}
\par\end{centering}

Supplementary Figure 6: Evolution of critical currents, in Device
C, oriented such that $\theta=\pi/2$. Data was taken with the top
gate voltage set to 0 V. \textbf{a)} The differential junction resistance,
measured with no applied DC current, evolves consistently with the
absence of BIA in the device. \textbf{b)} The critical currents in
the interference side lobes adjacent to the central lobe decay as
the parallel field $B_{x}$ increases from zero. The region of the
critical current measurement is outlined in red in (a). Each side
lobe is labeled in white. \textbf{c)} The extracted maximum critical
current for each side lobe. Within each lobe, the critical current
decreases from its maximum value at $B_{x}=0$ T.
\end{figure}
\begin{figure}[p]
\begin{centering}
\includegraphics[scale=0.35]{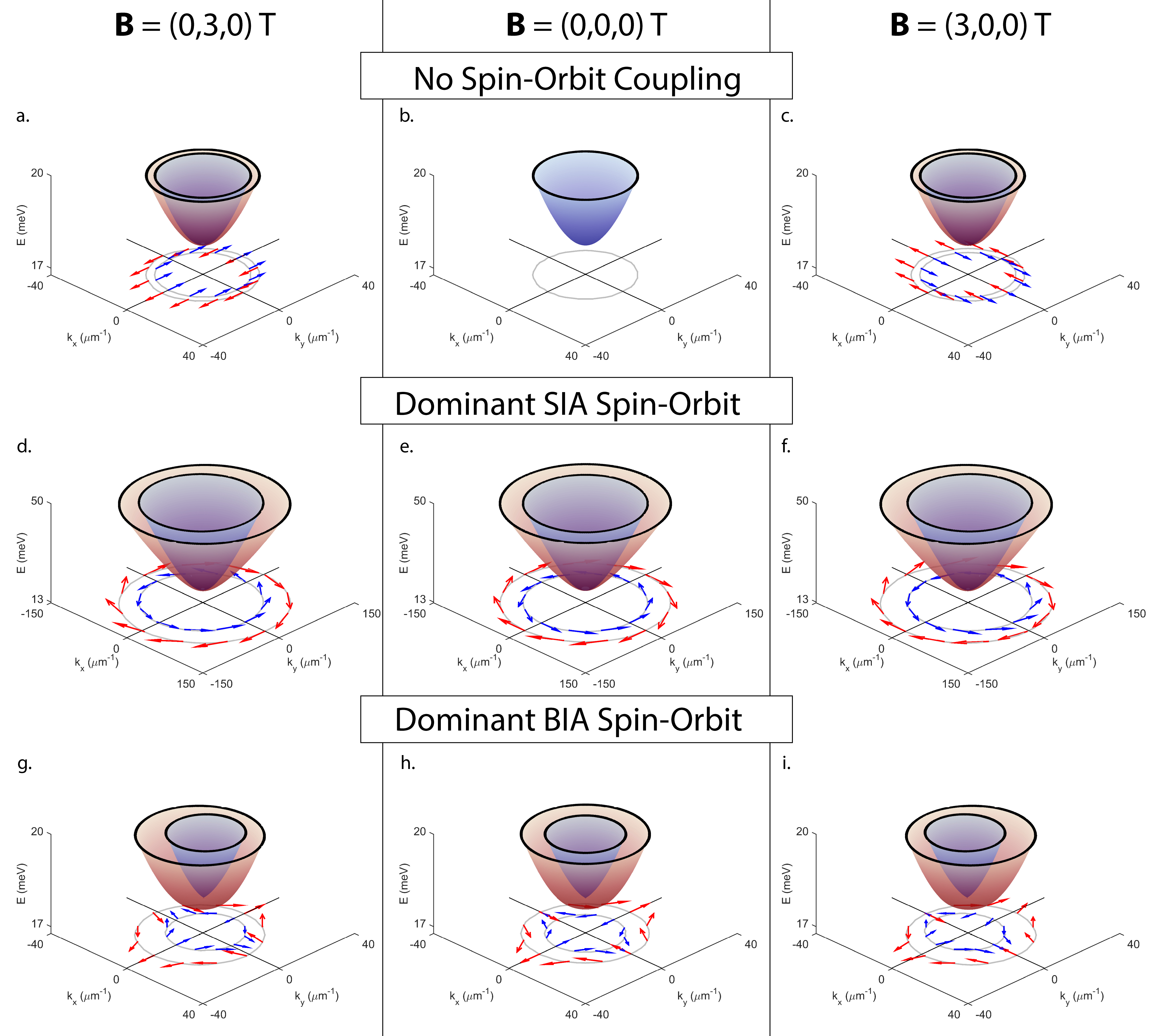}
\par\end{centering}

Supplementary Figure 7: Modeling of the conduction band structure
under various spin-orbit and parallel magnetic field conditions. \textbf{a-c)}
When spin-orbit coupling is absent, the addition of a 3 T parallel
magnetic field simply polarizes spins in a direction determined by
the sign of the in-plane g-factor, $g_{||}$. With no exernal magnetic
field, the two spin bands are degenerate (b). \textbf{d-f)} Dominant
structural inversion asymmetry causes axially symmetric spin-splitting
to occur at finite momentum (e). This type of spin-orbit coupling
causes the bands to shift orthogonally to the external magnetic field
(d, f), leading to nonzero Cooper pair momentum in the shift direction.
\textbf{g-i)} Dominant bulk inversion asymmetry with $\theta^{th}=0,\pi/2$
acts oppositely to structural inversion asymmetry, so that external
magnetic fields cause bands to shift parallel to the external magnetic
field. This type of shift is not observed experimentally in our devices. 
\end{figure}
}

\date{\noindent 
\begin{figure}[p]
\begin{centering}
\includegraphics[scale=0.35]{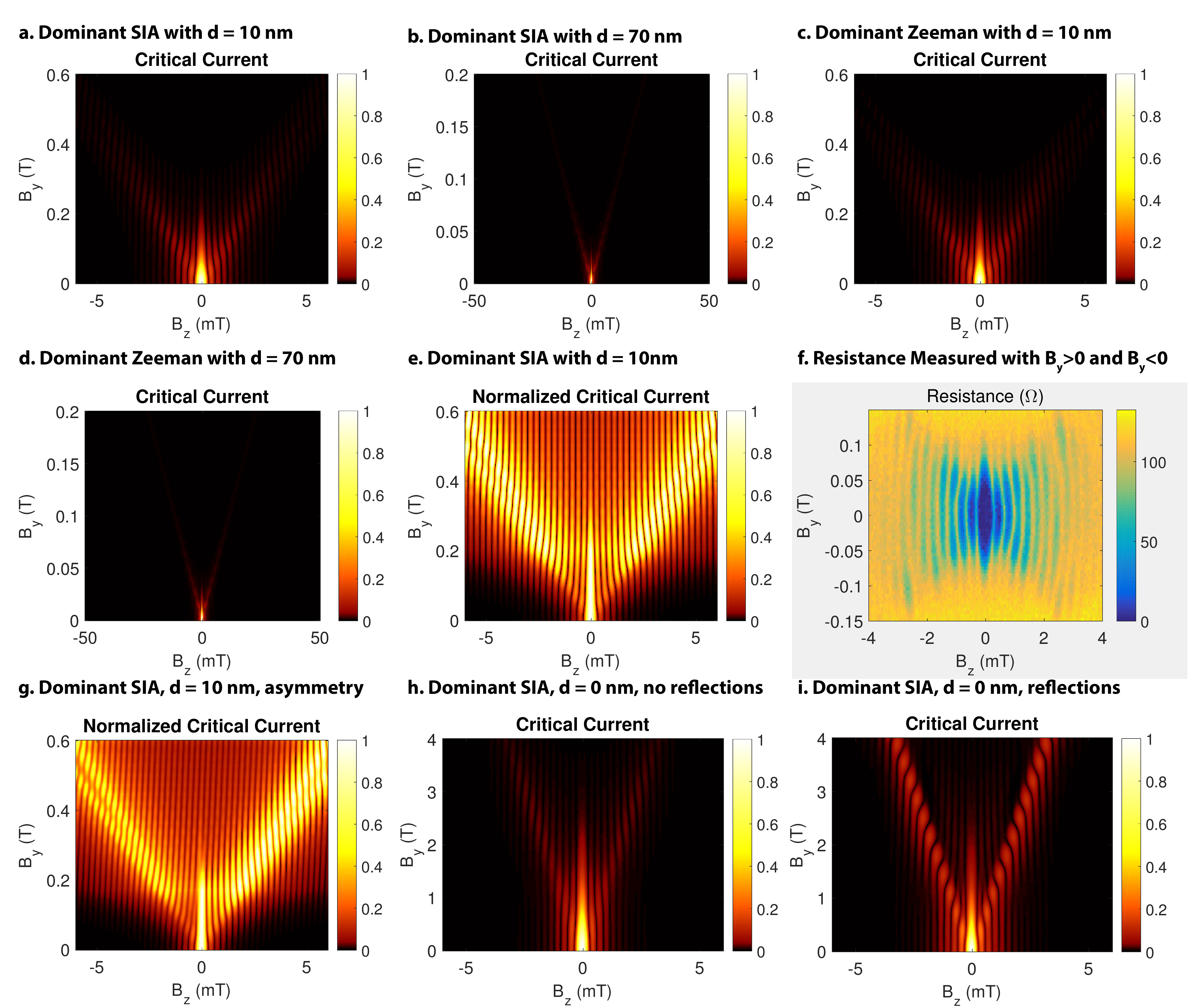}
\par\end{centering}

Supplementary Figure 8: Modeling of the critical current as the perpendicular
magnetic field $B_{z}$ (generating flux quanta) and the parallel
magnetic field $B_{y}$ (generating Cooper pair momentum) are varied.
\textbf{a)} With dominant SIA and the height difference $d$ between
the leads and the quantum well set to 10 nm, the interference evolves
consistently with measurements of differential resistance on aluminum
devices (devices A-D). \textbf{b)} Increasing $d$ to 70 nm decreases
the slope of each arm of the interference pattern consistently with
the measurement of a device with thickner niobium leads (device E).
\textbf{c, d)} Eliminating spin-orbit coupling leads to a similar
picture for both values of $d$, highlighting the overwhelming extrinsic
nature of the pairing momentum induced when the parallel field is
applied in the $y$ direction. \textbf{e)} Normalizing the critical
current at each value of $B_{y}$ reveals additional features weakly
present in the interference with SIA and $d=10$ nm. Since the extrinsic
wavevector $q_{y}$ adds and subtracts with the wavevector $\Delta k$
induced due to SIA, two slopes are in principle found in each arm
of the interference pattern. However, superconductivity weakens to
the extent that such splitting cannot be conclusively observed in
our devices. \textbf{f) }The resistance of Device F, measured as both
$B_{y}$ and $B_{z}$ are tuned to positive and negative values. The
measured resistance is observed to be symmetric under inversion of
both $B_{y}$ and $B_{z}$, as expected from time-reversal symmetry.
Under inversion of either $B_{y}$ or $B_{z}$, however, the resistance
is asymmetric. \textbf{g)} Modeling asymmetry in the lengths of superconducting
electrodes leads to asymmetry in the interference with respect to
inversion of $B_{z}$. Plotted here are the expected critical currents
for a device with 4 microns and 4.5 microns as the interfacial lengths.
\textbf{h) }The expected evolution of interference upon increasing
$B_{y}$ assuming dominant SIA and with $d=0$. \textbf{i)} Including
specular reflections at the mesa ends, assuming the presence of a
steep confining potential which does not flip spins upon reflection,
quantitatively modifies the interference evolution. However, qualitatively
the behavior remains unchanged.
\end{figure}
}
\end{document}